\documentclass[11pt]{article}
\usepackage[margin=1in]{geometry}
\usepackage[T1]{fontenc}
\usepackage{lmodern,microtype,amsmath,amssymb,amsthm,mathtools,bm,mathrsfs}
\usepackage{booktabs,array,enumitem,needspace}
\usepackage[colorlinks=true,linkcolor=blue,citecolor=blue,urlcolor=blue]{hyperref}
\hypersetup{pdftitle={Crossed Light OPEs and the Four-Gluon Spectral Bootstrap},
pdfsubject={Both crossed light blocks and the tree four-gluon spectral bootstrap}}
\allowdisplaybreaks[2]
\newtheorem{theorem}{Theorem}
\newtheorem{proposition}{Proposition}
\newcommand{\dd}{\mathrm d}
\newcommand{\ip}{\mathrm i}
\newcommand{\eps}{\epsilon}
\newcommand{\bz}{\bar z}
\newcommand{\cL}{\mathcal L}

\newcommand{\cB}{\mathcal B}
\newcommand{\cC}{\mathcal C}
\newcommand{\cG}{\mathcal G}
\newcommand{\cR}{\mathcal R}
\newcommand{\cA}{\mathcal A}
\newcommand{\Aplus}{\mathsf A^+}
\newcommand{\Aminus}{\mathsf A^-}

\newcommand{\nn}{\nonumber}
\title{Light-Transformed Carrollian OPEs and the\\Four-Gluon Spectral Bootstrap}
\author{%
  Sourish Banerjee$^{1,2}$ \qquad Rudranil Basu$^{1}$\\[0.8em]
  {\small $^{1}$Department of Physics, Birla Institute of Technology
  and Science, Pilani,}\\
  {\small K K Birla Goa Campus, Zuarinagar, Sancoale,
  Goa 403726, India}\\[0.5em]
  {\small $^{2}$Indian Institute of Technology Ropar,
  Punjab 140001, India}\\[0.8em]
  {\small E-mail:
  \href{mailto:p20210001@goa.bits-pilani.ac.in}
       {\texttt{p20210001@goa.bits-pilani.ac.in}},
  \quad
  \href{mailto:rudranilb@goa.bits-pilani.ac.in}
       {\texttt{rudranilb@goa.bits-pilani.ac.in}}}
}

\date{}
\begin{document}
\maketitle
\begin{abstract}
Scattering amplitudes of massless particles can be rewritten as
correlation functions of light-transformed operators at null infinity, on which the Carrollian symmetries act. We ask how much of the four-gluon tree amplitude is determined by the operator product expansion of these operators. When two gluons become collinear, the exchanged gluon and all of its descendants combine into a single function of the null time of $\mathcal{I}^+$ and angular separations of the pair on the celestial sphere. We show that this function is fixed by the Ward identities, which reduce to Kummer's differential equation, and by the signs of the energies, which select one combination of its two solutions. We then expand the same amplitude in massive two-particle states of the physical scattering channel and find that the collinear data control the expansion at large spin. Each of the two collinear poles of the amplitude determines one of the two infinite sequences of partial-wave coefficients in closed form. Once both poles are imposed, crossing symmetry and the scaling of the correlator in Carroll time fix the tree spectrum uniquely among amplitudes with only the physical collinear poles, up to a single subtraction constant at the collinear boundary.
\end{abstract}
\setcounter{tocdepth}{1}
\begingroup
\small
\setlength{\parskip}{0pt}
\tableofcontents
\endgroup
\clearpage

\section{Introduction}
\label{sec:introduction}

Scattering amplitudes of massless particles in flat space can be
rewritten as correlation functions of operators living at null
infinity \cite{Pasterski:2016qvg}. In the celestial formulation these operators are labelled
by a point on the sphere at infinity and a conformal weight \cite{Pasterski:2017kqt}. In the 
Carrollian formulation they also carry a retarded time label, and the
symmetries of flat space act on them as Carrollian conformal
symmetries \cite{BagchiEtAl, Duval:2014uva}. Reviews of celestial amplitudes and their relation to asymptotic
symmetries can be found in
\cite{Pasterski:2021rjz,Raclariu:2021zjz,Donnay:2023mrd}. The hope behind both formulations is a bootstrap for
scattering. If the operators at null infinity obey an operator
product expansion, and if the four-point correlator must be the same
whichever pair of operators is expanded first, then symmetry and
consistency alone might determine the amplitude. See \cite{Kruczenski:2022lot} for an equivalent approach named S-Matrix bootstrap.

Part of this programme is well understood. When two massless
operators approach each other on the sphere, the leading term of
their operator product is the collinear limit of the amplitude, and
symmetry fixes it almost completely \cite{PRSY}. The four-point problem asks
for more than the leading term. It needs the full contribution of
each exchanged operator together with all of its descendants, it
needs a basis of intermediate states with a definite normalization,
and it needs the crossing equation that relates the two ways of
pairing the external operators. One distinction turns out to be
essential. When two external operators have opposite energy signs,
their collinear limit is dominated by a single massless gluon, and
the infinite expansion contains the time and angular descendants
of that exchanged gluon. When the two operators have equal energy signs, their
total momentum is timelike, and the infinite expansion is a
continuum of massive two-particle states \cite{NguyenSalzer}. Both expansions are
infinite, but they describe different representations of the
Poincar\'e group, and a four-point bootstrap must keep them apart.

We work with finite-time light operators, which are well suited to
this distinction. The modified Mellin transform keeps the retarded
time as the variable conjugate to energy, so that time derivatives
produce translation descendants and the scaling of the correlator in
time records its energy dimension \cite{BanerjeeNullInfinity, Banerjee:2019prz}. For the light transform, we treat the two celestial coordinates as independent after analytic continuation. We integrate over one while holding the other and the retarded time fixed. The untransformed
four-point correlator is supported on a delta function in the
celestial coordinates, and the light integrations turn it into an
ordinary function on suitable coordinate regions
\cite{SharmaLight,DeEtAl,Hu:2022syq,Jorge-Diaz:2022dmy}. The construction combines the celestial
four-point kinematics of Pasterski, Shao and Strominger \cite{PSS}
with the Carrollian description of scattering developed in
\cite{DFHRPerspective,BagchiEtAl,DFHRBridge,MasonRuzziconiSrikant,Bagchi:2023fbj,BagchiDhivakarDuttaCarroll, Nguyen:2023miw}. Carrollian amplitudes can also be obtained from suitable flat limits of AdS correlators \cite{Alday:2024yyj, Lipstein:2025jfj}. A separate approach of relating the boundary correlators and scattering can be formulated through a generating functional as in \cite{Kraus:2024gso, Jain:2023fxc}.

In this paper we carry out the four-point analysis for the
alternating-helicity four-gluon tree amplitude. We first derive the
complete contribution of an exchanged massless gluon to the product
of two light operators, for both kinds of collinear pair that occur,
a pair of opposite helicities and a pair of equal helicities. We
then expand the same amplitude in the massive two-particle states of
the physical scattering channel and show that the two collinear
contributions are recovered from this expansion. Finally we show
that, once both collinear contributions are imposed, crossing
symmetry and the scaling of the correlator in Carroll time determine
the tree spectrum uniquely, which is indicative of the fact that Carrollian CFTs are natural duals to bulk quantum field theories \cite{Bagchi:2025vri}.

The two collinear pairs leave different fingerprints on the
spectrum. The opposite-helicity pair \((3^+,2^-)\) produces the pole
at \(u_{\!M}=0\), where \(u_{\!M}\) denotes the Mandelstam invariant
(the letter \(u\) is reserved for Carroll time), and an alternating
sequence of partial-wave coefficients built from harmonic numbers.
The same-helicity pair \((3^+,1^+)\) produces the pole at \(t=0\) and
a non-alternating sequence proportional to
\([(J-1)J(J+1)(J+2)]^{-1}\), where \(J\) is the spin of the
two-particle state. The second
pole is easy to miss, because on the physical angular interval it is
cancelled by a zero in the spinor numerator. Matching both
pairs fixes the entire infinite-spin part of the spectrum, including
every subleading inverse power of the spin.

Our work builds on several lines of research, which we now describe.
The symmetry constraints on the leading celestial operator product
have been obtained in complementary ways. Pate, Raclariu, Strominger
and Yuan fix the leading gluon and graviton coefficients through
recursion relations that follow from soft theorems \cite{PRSY}.
Banerjee and Ghosh derive the gluon operator product from celestial
current algebras \cite{BanerjeeGhosh}, and the gravitational
operator product is organized by BMS and current-algebra descendants
in ~\cite{BanerjeeGhoshGonzo,BanerjeeGhoshPaul}. Conformally
soft limits give another route to the same symmetry generators
\cite{AdamoMasonSharma,Fan:2019emx,Donnay:2018neh}. Guevara et al constructed towers of soft
currents \cite{GHPS}, whereas Strominger identified the \(w_{1+\infty}\) structure of the soft graviton algebra and its action on the soft gauge currents \cite{StromingerW}.
Himwich, Pate and Singh obtain the leading holomorphic operator
product for arbitrary spins and cubic couplings from Poincar\'e
symmetry alone \cite{HPS}. These algebras also have a twistor-space
description \cite{AMSTwistor}, including an explicit action on
Carrollian fields \cite{DFHTwistor}. On the other hand, a concrete relation between Carrollian current Ward identities and soft theorems and collinear limits has been established in \cite{Ruzziconi:2024kzo}.

Integral representations of scalar celestial OPE blocks, including
shadow and light-transformed contributions, were studied in
\cite{Guevara:2021tvr}. At tree level, the full collinear expansion
of maximally helicity violating gluon and graviton amplitudes can
also be obtained from on-shell recursion \cite{Ren:2023trv}.

The object we start from is the chiral Carrollian fusion kernel of
\cite{NguyenSalzer}, which contains the exchanged gluon with its time
and angular descendants and whose integration range is selected by
the energy signs. We study the image of this
kernel under the light transform. The light transform and its
relation to the shadow transform have a general Lorentzian
formulation \cite{KravchukSimmonsDuffin}, and their celestial and
finite-time Carrollian versions are developed in \cite{SharmaLight,BBBLight,Banerjee:2025oyu}. Operator products of
null-integrated operators have also been constructed in general
Lorentzian conformal field theories \cite{Kologlu:2019mfz}. In our setting the
Ward identities, written in the variable conjugate to the energy
fraction, become a second-order differential equation. The
integration range inherited from the energy signs then selects one
combination of its two solutions, and we check the normalization of
this combination against the amplitude. We must emphasize on the fact that it is only because we make the switch to the light-transformed basis that we obtain this second-order differential equation. Both the solutions turn out to be crucial to understand the four-particle scattering amplitude.

For the two-particle states we use the Jacob-Wick basis, which is
orthogonal and labelled by invariant mass and rest-frame spin
\cite{JacobWick}. Law and Zlotnikov construct the corresponding
celestial partial waves and discuss the divergence that arises when
a massless particle is exchanged in a collinear configuration\cite{LawZlotnikov}. Decompositions of four-gluon
correlators into Lorentz conformal blocks give a complementary
organization of the same data \cite{FanI,FanII,Nandan:2019jas}, as does the
analysis of discrete and light-ray contributions in scalar exchange
\cite{AtanasovEtAl}. Our relation between the spectrum and the
collinear contributions uses the universal collinear splitting of
gluon amplitudes \cite{BDDK} and keeps the full time dependence of
the exchanged family.

Crossing-symmetric dispersion relations offer a related way to
organize scattering data \cite{SinhaZahed}. Their celestial form and
the locality constraints they imply are studied in \cite{GhoshRamanSinha}, and Liu and Ma reconstruct tree MHV
amplitudes from a crossing-symmetric celestial operator product
\cite{LiuMa}. The question we address is different in emphasis. We
ask how the complete finite-time contribution of a massless exchange
is related to normalized two-particle spectral coefficients. This
connects the massless operator product to the multiparticle sectors
studied in \cite{GuevaraHuPasterski,KulpPasterski}, and it
provides the tree-level data on which the loop-level associativity
problem of Costello and Paquette is built \cite{CostelloPaquette}.

Three qualifications apply to all our statements. We consider
massless exchange at nonzero frequency, and we define light
transforms as boundary values along the contours specified in
Appendix~\ref{app:cycles}. The relation between spectrum and
collinear contributions is obtained from the analytically continued
and resummed partial-wave expansion, at its crossed collinear pole.
Massless exchange also requires a subtraction prescription for the collinear divergence, and we hold this prescription fixed when comparing
channels. Within the class of rational tree amplitudes specified in
Section \ref{sec:spectralunique}, these qualifications are all that
is needed for the two collinear contributions, crossing and time
scaling to determine the spectrum.

\subsection{Summary of results}

For a pair of opposite-helicity light operators, the Ward identities
reduce the contribution of the exchanged gluon to Kummer's equation
in a single variable. The equation has two solutions, the expected
tower of time descendants and a second tower related to it by a
reflection of the weights, and the integration range fixed by the
energy signs selects a definite combination of the two, a Tricomi
function. For a pair of same-helicity operators the same steps give
Kummer's equation with a second parameter shifted by one, and with an
argument that contains both the time separation and the angular
separation of the pair. We then show that summing the normalized
two-particle partial waves and taking the discontinuity at a
collinear pole reproduces the corresponding complete contribution.
The opposite-helicity pole gives the alternating sequence of
harmonic numbers, and the same-helicity pole gives the
non-alternating rational sequence. Once both poles are matched, the
remaining freedom in a rational tree amplitude occupies finitely many
spins in each physical channel, and crossing together with the time
scaling of Yang--Mills theory removes it.

Section~\ref{sec:conventions} fixes the transforms and the symmetry
generators. Section~\ref{sec:chiral} derives the two collinear
contributions from the Ward identities. Section~\ref{sec:partialwaves}
constructs the normalized two-particle spectrum.
Section~\ref{sec:crossing} relates the two, writes the crossing
equation, and proves the uniqueness result.
Section~\ref{sec:discussion} explains what the light basis
contributes and describes the loop-level problem that follows. The
three appendices contain the fusion integrals, the projection
identities and reproducible checks.
\section{Light operators and their symmetry action}
\label{sec:conventions}

We fix the transform conventions and the symmetry action used in the
block construction. The induced translations keep track of energy
insertions, while time translation symmetry will later restrict the mass dependence of the spectral coefficients.

We use the metric \((-+++)\), signed all-outgoing momenta
\begin{equation}
 p_i^\mu=\eps_i\omega_i q^\mu(z_i,\bz_i),\qquad
 q^\mu=(1+z\bz,z+\bz,-\ip(z-\bz),1-z\bz),\qquad \omega_i>0.
\end{equation}
Here \(\eps_i=+1\) labels an outgoing particle and \(\eps_i=-1\)
an incoming one. We write \(z_{ij}=z_i-z_j\) and
\(\bz_{ij}=\bz_i-\bz_j\). On real Lorentzian kinematics,
\(-(p_i+p_j)^2=4\eps_i\eps_j\omega_i\omega_j|z_{ij}|^2\).
Thus a generic equal-sign pair of momenta has timelike total momentum and resolves
into massive two-particle states; we call this the \textit{physical
channel}. An opposite-sign pair has spacelike total momentum, and its only
singularity is the collinear pole, where a massless gluon is exchanged; we
call this the \textit{crossed channel}. To evaluate the light
integrals, we choose a connected region of coordinates in which
the integration contours can be varied continuously without crossing
a singularity \cite{Banerjee:2022hgc}. We call such a region a \emph{chamber}. In a real coordinate chart, it is specified by inequalities between the coordinates, such as \(z_1<z_2\), together with consistent branch choices for the complex powers. We will state the relevant inequalities when each light integral is introduced.

The modified Mellin fields \cite{BanerjeeNullInfinity, Banerjee:2019prz} and light transforms are
\begin{align}
 O_{\Delta,\ell}^{\eps}(u,z,\bz)
 &=\int_0^\infty\dd\omega\,\omega^{\Delta-1}
 e^{-\ip\eps\omega u}a_{\ell}^{\eps}(\omega,z,\bz),&
 \partial_u O_{\Delta,\ell}^{\eps}&=-\ip\eps O_{\Delta+1,\ell}^{\eps},
 \label{eq:mm}\\
 \cL_h^+[O]&=\int_{\Gamma_z}\dd w\,(w-z)^{2h-2}O(u,w,\bz),&
 \cL_{\bar h}^-[O]&=\int_{\Gamma_{\bz}}\dd\bar w\,
 (\bar w-\bz)^{2\bar h-2}O(u,z,\bar w),\nn\\
 h&=\frac{\Delta+\ell}{2},\qquad
 \bar h=\frac{\Delta-\ell}{2},&
 \Aplus_\Delta&=\cL^+[O_{\Delta,+1}],\quad
 \Aminus_\Delta=\cL^-[O_{\Delta,-1}].
 \label{eq:light}
\end{align}
Here \(\Delta\) is the scaling dimension of the conformal
Carroll primary, and \(\ell\) is its spin, equal to the helicity
of the associated massless particle.$a_{\ell}^{\eps}(\omega,z,\bz)$ is the bulk momentum mode. For complex weight, the oriented real-line integral is a boundary
value from above or below the kernel's cut,
\((w-z+\ip\varsigma0)^{2h-2}\), with \(\varsigma=\pm1\).
Both contours run from \(-\infty\) to \(+\infty\). The two boundary
choices are supported on opposite halves of the angular Fourier axis:
writing \(\xi\) for the wave number conjugate to the light coordinate
(it is independent of the energy \(\omega\), which is conjugate to
Carroll time, and is not restricted by \(\omega>0\)), the kernel acts
on Fourier modes \(e^{\ip\xi w}\) as the multiplier
\(\propto\Theta(-\varsigma\xi)|\xi|^{1-2h}\) of
\eqref{eq:lightmultiplier}, so that \(\varsigma=+1\) retains \(\xi<0\)
and \(\varsigma=-1\) retains \(\xi>0\), for either energy sign \(\eps\).
We therefore keep both choices on each leg, so that the
nonzero-frequency part of the original angular wave packet can be
reconstructed from its light transforms. We suppress these labels in componentwise block
formulas. The boundary choices determine the phases and the continuation
of the block arguments. We will discuss these choices when evaluating the light integrals. The exchanged families and their spectral coefficients are the same for both choices.

The \(\omega_i\) integrals in \eqref{eq:mm} are initially evaluated
at complex Carroll times,
\begin{equation}
 u_i=u_i^{\rm R}-\ip\eps_i\eta_i,\qquad \eta_i>0.
 \label{eq:timeboundary}
\end{equation}
With this choice, the Fourier factor becomes
\[
 e^{-\ip\eps_i\omega_i u_i}
 =e^{-\ip\eps_i\omega_i u_i^{\rm R}}e^{-\eta_i\omega_i}.
\]
The factor \(e^{-\eta_i\omega_i}\) damps the integrand at large
energy. After integration, we obtain the correlator at real times
by taking \(\eta_i\to0^+\), while retaining the Feynman prescription
of the amplitude.

We denote by \(\mathrm{MM}\) the product of the four modified
Mellin transforms in \eqref{eq:mm}, one for each external leg.
Applying the four light transforms afterwards gives the complete
transform,
\[
 \mathsf T=\cL_1\cL_2\cL_3\cL_4\,\mathrm{MM},
\]
where \(\cL_i\) is the light transform appropriate to the helicity
of leg \(i\), as defined in \eqref{eq:light}. For a four-gluon
amplitude \(\cA\), we define the light-transformed four-point
correlator by
\begin{equation}
 \cG^L_{\boldsymbol\Delta}(u_i,z_i,\bz_i)
 =
 \mathsf T\!\left[
 (2\pi)^4\delta^{(4)}\!\left(\sum_{i=1}^4p_i\right)\cA
 \right],
 \qquad
 \boldsymbol\Delta=(\Delta_1,\Delta_2,\Delta_3,\Delta_4).
 \label{eq:lightcorrelator}
\end{equation}
The energy-sign, helicity and colour-ordering labels are left
implicit. The common
factor \(\ip\) relating \(\cA\) to the connected \(S\)-matrix element is
stripped. We work with colour-ordered amplitudes. Here the ordering specifies
the cyclic sequence of external gluons in the trace of gauge-group
generators multiplying each partial amplitude. We use the orderings
\((1,2,3,4)\) and \((1,3,2,4)\), with external helicities
\((1^+,2^-,3^+,4^-)\). Each partial amplitude includes the coupling
factor \(g^2\).
The helicity labels on an incoming all-outgoing leg are the negatives
of its physical incoming helicities.

The untransformed global generators, with conventional factors of
\(\ip\) suppressed, are
\begin{align}
 P_{00}&=\partial_u,&P_{10}&=z\partial_u,&
 P_{01}&=\bz\partial_u,&P_{11}&=z\bz\partial_u,\label{eq:P}\\
 L_{-1}&=\partial_z,&
 L_0&=z\partial_z+h+\tfrac12u\partial_u,&
 L_1&=z^2\partial_z+2hz+zu\partial_u.
 \label{eq:L}
\end{align}
There are also the barred Lorentz generators. We obtain all
light-basis generators by induction,
\begin{equation}
 \widehat G\,\cL[O]=\cL[G O].
 \label{eq:induced}
\end{equation}
For a plus field, put \(d=1-2h\). Inserting \eqref{eq:mm} gives
\(\widehat P_{00}\Aplus_\Delta=-\ip\eps\cL_h^+[O_{\Delta+1,+1}]\).
Since \(\partial_z(w-z)^{2h-1}=d(w-z)^{2h-2}\), differentiation
of the kernel, or integration by parts in \(w\), gives
\(\partial_z\Aplus_{\Delta+1}=d\cL_h^+[O_{\Delta+1,+1}]\).
The induced translations are therefore
\begin{align}
 \widehat P_{00}\Aplus_\Delta
 &=-\ip\eps d^{-1}\partial_z\Aplus_{\Delta+1},&
 \widehat P_{10}\Aplus_\Delta
 &=-\ip\eps(zd^{-1}\partial_z+1)\Aplus_{\Delta+1},\nn\\
 \widehat P_{01}&=\bz\widehat P_{00},&
 \widehat P_{11}&=\bz\widehat P_{10}.
 \label{eq:lightP}
\end{align}
For a minus field interchange \(z,h\) with \(\bz,\bar h\) and
\(P_{10}\) with \(P_{01}\). The Lorentz action on a plus field includes
\begin{align}
 \widehat L_{-1}\Aplus_\Delta&=\partial_z\Aplus_\Delta,\nn\\
 \widehat L_0\Aplus_\Delta
 &=(z\partial_z+1-h)\Aplus_\Delta
       +\tfrac12u\widehat P_{00}\Aplus_\Delta,\nn\\
 \widehat L_1\Aplus_\Delta
 &=(z^2\partial_z+2(1-h)z)\Aplus_\Delta
       +u\widehat P_{10}\Aplus_\Delta.
 \label{eq:lightL}
\end{align}
These operators are first order in the celestial coordinate.
Translations shift \(\Delta\) by one because they insert one power
of energy. The angular Lorentz terms preserve \(\Delta\), while
their time-dependent terms contain the same shifted translations.
The barred action follows from \eqref{eq:L} with the untransformed
barred weight \cite{BBBLight}.

The energy dependence of the amplitude determines how the
correlator changes under a common rescaling of the Carroll times.
Suppose \(\cA\) is homogeneous of degree \(k\) under a common
rescaling of all four energies, with the angular coordinates held
fixed. In the definition \eqref{eq:lightcorrelator}, changing
variables to \(\omega_i'=\lambda\omega_i\), with \(\lambda>0\),
gives a factor \(\lambda^{-\sum_i\Delta_i}\) from the Mellin
measures, \(\lambda^4\) from the momentum-conservation delta
function, and \(\lambda^{-k}\) from the amplitude. The light
kernels are independent of the energies. Therefore,
\begin{equation}
 \cG^L_{\boldsymbol\Delta}(\lambda u_i,z_i,\bz_i)
 =\lambda^{-(\Lambda+k)}
 \cG^L_{\boldsymbol\Delta}(u_i,z_i,\bz_i),
 \qquad \Lambda=\sum_i\Delta_i-4.
 \label{eq:homogeneity}
\end{equation}
Here all four times are rescaled together, including their
imaginary parts in \eqref{eq:timeboundary}. The prescription for
treating collinear singularities is held fixed and introduces
no energy scale. Consequently, the same relation holds after
taking the real-time limit \(\eta_i\to0^+\).

Tree Yang-Mills amplitudes have \(k=0\), so their light
correlators scale as \(\lambda^{-\Lambda}\). This relation
expresses the energy homogeneity of the amplitude directly
in Carroll-time variables and provides a constraint beyond
Lorentz covariance.

\section{Chiral light blocks from symmetry}
\label{sec:chiral}
In this section we construct the building blocks that will later be
matched to the crossed collinear channels, namely the contribution of a single massless exchanged operator, together with all of its Carroll-time and
angular descendants, to the product of two light-transformed gluon
operators. We call such a contribution a \emph{crossed light block}, or
simply a light block. The main
result is that for a specified exchanged family, the Ward identities determine a two-dimensional space of reduced
blocks. Once the two external operators are light transformed, the
Ward identity for the exchanged family becomes a second-order
ordinary differential equation, Kummer's equation, and therefore has
two independent solutions. Which combination of the two appears in the
physical block is decided by the range of the energy-fraction integral,
which in turn is fixed by the energy signs of the two external
operators. The answer is a Tricomi function.

Before the light transforms, the object we start from \footnote{The
roles of \(z\) and \(\bz\) are interchanged to match the branch used
here.} is the chiral
Carrollian fusion kernel \cite{NguyenSalzer}. We rederive the kernel from the Ward
identities with its density left unknown, because we need the
identities in a form that can be followed through the light transforms.
We first treat a pair of opposite-helicity operators, then a pair of
the same helicity. The two cases differ in which light transforms are holomorphic, and this difference will later account for their different large-spin tails in the timelike spectrum.

Throughout, every block carries one overall constant
\(N_g^\Gamma=(g/2)e^{\ip\phi_\Gamma}\), depending upon the light transform contour $\Gamma$. The prefactor \(g/2\) is the
cubic Yang-Mills normalization and is fixed in Section~\ref{sec:bridge}
by comparison with the four-gluon amplitude. Its phase \(\phi_\Gamma\), in general complex,
depends on the boundary prescription chosen for the light
contours. This is discussed in Appendix~\ref{app:cycles}, and vanishes for
the untransformed kernel. Since the phase plays no role in the
arguments below, we usually write \(N_g\) and suppress the label.

\subsection{Opposite-helicity pair: from the Ward identity to Kummer's equation}

Consider two external operators of opposite helicity, leg~1 with
helicity \(+\) and leg~2 with helicity \(-\), and opposite energy
signs, \(\eps_1=-\eps_2\). As explained in Section~\ref{sec:conventions},
such a pair has spacelike total momentum and becomes collinear at the
boundary of its crossed channel. In the collinear limit a single
massless gluon is exchanged. For this helicity assignment it has
helicity \(-\) and energy sign \(\eps_p=\eps_2\). Its scaling dimension is
\(p=\Delta_1+\Delta_2-1\), the familiar weight of the leading celestial
gluon OPE. It is convenient to introduce
\begin{equation}
 a=\Delta_1-1,\qquad b=\Delta_2,\qquad
 p=\Delta_1+\Delta_2-1=a+b,\qquad
 v=u_1-u_2,\quad x=\bz_2-\bz_1.
 \label{eq:abp}
\end{equation}
Here \(v\) is the Carroll-time separation and \(x\) the separation
of the antiholomorphic positions of the two external operators.

\paragraph{Shape of the fusion kernel.}
We first explain why the kernel has the form it has before deriving it. Three facts about the collinear limit are of importance.

We write \(\lambda_i\) and \(\widetilde\lambda_i\) for
the two momentum spinors associated with the null momentum
of leg \(i\). The brackets \(\langle ij\rangle\) and \([ij]\)
are their antisymmetric contractions,
\[
 \langle ij\rangle
 =\epsilon_{\alpha\beta}\lambda_i^\alpha\lambda_j^\beta,
 \qquad
 [ij]=\epsilon_{\dot\alpha\dot\beta}
 \widetilde\lambda_i^{\dot\alpha}
 \widetilde\lambda_j^{\dot\beta},
\]
where the antisymmetric spinor tensor is normalized by
\(\epsilon_{12}=1\).

First, the collinear singularity of a gluon amplitude is a
holomorphic pole, \(1/\langle12\rangle\propto1/z_{12}\), with
\(\bz_{12}\) held fixed. In the Carrollian setting the operator
product is extracted from the discontinuity of the correlator across
this pole, and the discontinuity of a simple pole is a delta function,
\(1/(z-\ip0)-1/(z+\ip0)=2\pi\ip\,\delta(z)\). The singular part of the
product is therefore supported on \(z_{12}=0\) at fixed \(\bz_{12}\), earning it the qualification \emph{chiral}. The more familiar
power-law behaviour \(z_{12}^{a}\) appears only after the light
transform of leg~1 is performed.

Second, in the collinear limit the two momenta add up to the
exchanged momentum, \(p_1+p_2=p_p\). Let
\(\tau=\eps_1\omega_1/(\eps_p\omega_p)\) be the signed energy fraction
carried by leg~1. Adding the two null momenta \(\eps_i\omega_i q(z_i,\bz_i)\)
and comparing with \(\eps_p\omega_p q(z_p,\bz_p)\) shows that, at
\(z_1=z_2\), the exchanged operator sits at the energy-weighted
position \(\bz_p=\tau\bz_1+(1-\tau)\bz_2=\bz_2+\tau\bz_{12}\). The same weighted
average appears in the time argument. The modified Mellin phases of the two
external operators combine as
\(e^{-\ip\eps_1\omega_1u_1-\ip\eps_2\omega_2u_2}
=e^{-\ip\eps_p\omega_p(u_2+\tau v)}\). Thus the exchanged operator is inserted at
\(Y(\tau)=(u_2+\tau v,\,z_2,\,\bz_2+\tau\bz_{12})\).

Third, the range of \(\tau\) is fixed by the energy signs. With
\(\eps_1=-\eps_2=-\eps_p\), one has \(\omega_1=-\tau\omega_p\) and
\(\omega_2=(1-\tau)\omega_p\), and positivity of both energies requires
\(\tau<0\). The energy fraction is therefore integrated over
\(\mathcal I=(-\infty,0)\).

We now combine these facts to write the contribution of the exchanged negative-helicity gluon to the OPE of the two original Carrollian operators,
\[
 O_{\Delta_1,+1}^{\eps_1}(u_1,z_1,\bz_1)\,
 O_{\Delta_2,-1}^{\eps_2}(u_2,z_2,\bz_2).
\]
The external light transforms have not yet been applied. The exchanged operator is \(O_p\equiv O_{p,-1}^{\eps_p}\), with \(p=\Delta_1+\Delta_2-1\) and \(\eps_p=\eps_2\). We retain its contribution and the descendants obtained by varying its insertion point along \(Y(\tau)\).

As described above, we represent the collinear pole by its discontinuity divided by \(2\pi\ip\), which replaces \(1/z_{12}\) by \(\delta(z_{12})\). We denote the resulting contribution by \(\mathbb K^C_\rho[O_p]\) and write
\begin{equation}
 \mathbb K^C_\rho[O_p]
 =N_g\delta(z_{12})\int_{\mathcal I}\dd \tau\,
 \rho(\tau)O_p(Y(\tau)),\qquad
 Y(\tau)=(u_2+\tau v,z_2,\bz_2+\tau\bz_{12}).
 \label{eq:unknownparent}
\end{equation}
with a density \(\rho(\tau)\) that remains to be determined. The superscript \(C\) denotes the Carrollian parent before
the external light transforms. Its light image will carry a
superscript \(L\).

Nothing more general is allowed by the symmetries. Let the time and
angular arguments of the exchanged operator carry independent
fractions, \(u_p=u_2+\tau v\) and \(\bz_p=\bz_2+\tau'\bz_{12}\), and let
the density depend on \(v\) and \(\bz_{12}\) as well. The dilatations
\(L_0\) and \(\bar L_0\) give \(h_1+h_2-h_p=1\), the extra unit being
supplied by \(\delta(z_{12})\), and \(\bar h_1+\bar h_2-\bar h_p=0\), so
the density cannot depend on either separation. Acting with
\(P_{01}=\bz\partial_u\) on the two external legs and on the exchanged
operator, with the common factor \(N_g\delta(z_{12})\rho(\tau)\) inert,
gives
\begin{equation}
 (\bz_1\partial_{u_1}+\bz_2\partial_{u_2})O_p(u_p,z_2,\bz_p)
 -\bz_p\partial_{u_p}O_p(u_p,z_2,\bz_p)
 =(\tau-\tau')\bz_{12}\,\partial_{u_p}O_p(u_p,z_2,\bz_p)=0,
 \label{eq:neweq}
\end{equation}
so for an exchanged mode of nonzero frequency and \(\bz_{12}\ne0\),
\(\tau'=\tau\): the same energy fraction fixes the time and the
antiholomorphic position of the exchanged operator. The remaining three
translations then hold automatically, because
\begin{equation}
 \tau f(z_1,\bz_1)+(1-\tau)f(z_2,\bz_2)
 =f(z_2,\bz_2+\tau\bz_{12}),\qquad
 f=1,z,\bz,z\bz,\quad z_1=z_2.
 \label{eq:translationfusion}
\end{equation}

\paragraph{The Ward identity for the density.}
It remains to impose the Lorentz generators, and only \(\bar L_1\)
gives a nontrivial condition. For any generator \(G\) we compare its
action on the two external operators with its action on the exchanged
one,
\[
 \mathcal W_G\mathbb K^C_\rho[O_p]
 =(G^{(1)}+G^{(2)})\mathbb K^C_\rho[O_p]
 -\mathbb K^C_\rho[G^{(p)}O_p].
\]
and the kernel is compatible with the symmetry precisely when
\(\mathcal W_G\) vanishes for every \(G\). For \(\bar L_1\) the barred
weights are \(2\bar h_1=a\), \(2\bar h_2=b+1\) and \(2\bar h_p=p+1\),
and a direct computation gives
\begin{equation}
 \mathcal W_{\bar L_1}\mathbb K^C_{\rho}[O_p]
 =N_g\delta(z_{12})\bz_{12}\int\dd \tau\,\rho(\tau)
 \left[\tau(1-\tau)\partial_\tau+a-(p+1)\tau\right]O_p(Y(\tau)).
 \label{eq:parentwardaction}
\end{equation}
The derivative \(\partial_\tau O_p(Y(\tau))=v\,\partial_uO_p+\bz_{12}\partial_{\bz}O_p\)
collects both the angular and the time dependence of the exchanged
operator. The time part is produced by the term \(\bz u\partial_u\) in
\(\bar L_1\). Since the bracket in \eqref{eq:parentwardaction} is a
first-order differential operator in \(\tau\) acting on \(O_p\), we can
integrate by parts and let it act on the density instead. Requiring
the result to vanish for an arbitrary exchanged field gives
\begin{equation}
 \partial_\tau[\tau(1-\tau)\rho]-[a-(p+1)\tau]\rho=0,
 \label{eq:rhoward}
\end{equation}
up to boundary terms at the ends of \(\mathcal I\), which are
discussed in Appendix~\ref{app:cycles}. Equation \eqref{eq:rhoward} is
the density constraint of \cite{NguyenSalzer}. Its solution is
elementary, but we deliberately do not solve it yet, as we want to see
what the light transforms do to the equation itself.

\paragraph{Light transforming the two external legs.}
We now apply \(\cL_1^+\) to leg~1 and \(\cL_2^-\) to leg~2. On leg~1
the light integral runs over the holomorphic position \(w\) against
the kernel \((w-z_1)^{a}\), and the delta function
\(\delta(w-z_2)\) simply evaluates it, giving
\((-z_{12}+\ip\varsigma_10)^{a}\). Starting in the region
\(z_{12}>0\), its boundary value is
\[
 (-z_{12}+\ip\varsigma_1 0)^a
 =e^{\ip\pi\varsigma_1a}z_{12}^a.
\]
We absorb the branch factor into
\(N_g^{\Gamma_1}=e^{\ip\pi\varsigma_1a}N_g^{\rm parent}\), where
\(N_g^{\rm parent}\) is the constant of the untransformed kernel
\eqref{eq:unknownparent}, fixed to \(g/2\) in Section~\ref{sec:bridge}.
This gives the power \(z_{12}^a\) announced above. Other coordinate
regions are reached by continuing the same branch prescription.

On leg~2 the light integral runs over the antiholomorphic position
\(\bar w_2\) against the kernel \((\bar w_2-\bz_2)^{b-1}\). Inside the
integrand the exchanged operator now sits at
\(\tau\bz_1+(1-\tau)\bar w_2\). It is convenient to use this fused position
itself as the integration variable, \(\bar w=\tau\bz_1+(1-\tau)\bar w_2\),
and to measure it from \(\bz_2\) by setting \(W=\bar w-\bz_2\). The
external kernel and the Jacobian of this change of variables then
combine into the explicit factors \((1-\tau)^{-b}(W+\tau x)^{b-1}\), which we
absorb together with \(\rho\) into a new density \(H\). The result is
\begin{align}
 \mathbb K^L_H[O_p]&=N_g z_{12}^a\int\dd\bar w\,\dd \tau\,
 H(\tau;W,x)O_{p,-1}^{\eps_p}(u_2+\tau v,z_2,\bar w),\nn\\
 H(\tau;W,x)&=(1-\tau)^{-b}(W+\tau x)^{b-1}\rho(\tau).
 \label{eq:unknownlight}
\end{align}
Thus \(\mathbb K^L_H=\cL_1^+\cL_2^-\mathbb K^C_\rho\). Note that the complex power in \(H\) retains the boundary prescription of the
leg-2 light kernel. The exchanged operator is now integrated over its antiholomorphic
position, so the block no longer depends on \(\bz_{12}\) explicitly;
the angular separation \(x\) enters only through the density.

Rewriting \eqref{eq:rhoward} in terms of \(H\) gives
\begin{equation}
 \mathfrak W_\tau H\equiv
 \left[\tau(W+\tau x)\partial_\tau-(a-1)W-(p-2)\tau x\right]H=0.
 \label{eq:lightdensityward}
\end{equation}
Computing the same difference directly with the induced Lorentz
generators \eqref{eq:lightL} gives, (one can verify that the Lorentz residual computed directly on the light-transformed kernel
\begin{equation}
 \mathcal W_{\bar L_1}\mathbb K^L_H[O_p]
 =N_g z_{12}^a\int\dd\bar w\,\dd \tau\,
 \frac{W+x}{W+\tau x}\mathfrak W_\tau H\,
 O_p(u_2+\tau v,z_2,\bar w),
 \label{eq:directlightward}
\end{equation}
again up to boundary terms, so \eqref{eq:lightdensityward} is the
Ward identity of the light block in its own right.

\paragraph{From a first-order to a second-order equation.}
The decisive step is to look at the Ward identity not in the energy
fraction \(\tau\) but in the variable conjugate to it. An exchanged mode
of energy \(\omega\) inserted at time \(u_2+\tau v\) carries the phase
\(e^{-\ip\eps_p\omega(u_2+\tau v)}
=e^{-\ip\eps_p\omega u_2}e^{-\varpi\tau}\), where
\begin{equation*}
 \varpi=\ip\eps_p\omega v
\end{equation*}
is, up to a phase, the product of the exchanged energy and the time
separation. The \(\tau\)-integral in \eqref{eq:unknownlight} is therefore
a Laplace transform of the density,
\(S(\varpi;W,x)=\int\dd \tau\,H\,e^{-\varpi\tau}\). Multiplying
\eqref{eq:lightdensityward} by \(e^{-\varpi\tau}\) and integrating over \(\tau\),
each power of \(\tau\) becomes a derivative with respect to \(\varpi\), and
one finds, for any density \(H\),
\begin{equation}
 \int\dd \tau\,e^{-\varpi\tau}\mathfrak W_\tau H
 =[\tau(W+\tau x)He^{-\varpi\tau}]_{\partial\mathcal I} +\left[\varpi x\partial_\varpi^2+(px-\varpi W)\partial_\varpi-aW\right]S.
 \label{eq:wardlaplace}
\end{equation}
The definition of \(H\) and the density constraint determine its
scaling weights. Specifically, \eqref{eq:unknownlight} and
\eqref{eq:rhoward} give
\begin{align*}
 (W\partial_W+x\partial_x)H=(b-1)H, \qquad
 (\tau\partial_\tau-x\partial_x)H=(a-1)H.
\end{align*}
Taking the Laplace transform and integrating the \(\tau\)-derivative
by parts, with vanishing boundary terms, gives
\begin{align*}
 (W\partial_W+x\partial_x)S =(b-1)S,\qquad
 (\varpi\partial_\varpi+x\partial_x)S=-aS.
\end{align*}
These two equations fix the prefactor and leave one independent
combination of variables,
\begin{equation*}
 S=x^{-a}W^{p-1}F(\chi),\qquad \chi=\frac{\varpi W}{x},
\end{equation*}
where the single variable \(\chi\) combines the exchanged energy, the
time separation and the ratio of the two angular separations. When
the boundary term in \eqref{eq:wardlaplace} vanishes, the Ward
identity becomes
\begin{equation}
 \boxed{\displaystyle
 \mathcal D_{a,p}F\equiv
 [\chi\partial_\chi^2+(p-\chi)\partial_\chi-a]F=0.}
 \label{eq:kummerode}
\end{equation}
This is Kummer's confluent hypergeometric equation \cite{Kummer+1837+228+242}. Its origin is easy to state in \eqref{eq:lightdensityward} the Lorentz generator acts on
the energy fraction through \(\tau(W+\tau x)\partial_\tau\), which is quadratic
in \(\tau\), and under the Laplace transform each power of \(\tau\) becomes
a derivative in \(\varpi\). A first-order equation with quadratic
coefficients in \(\tau\) therefore turns into a second-order equation in
\(\chi\), and a second-order equation has two independent solutions. The two local solutions therefore follow from the Ward identities before the fusion integral is evaluated. The next subsection selects their physical combination and fixes its normalization, absorbing the remaining coordinate-independent branch factor into \(N_g\).
\subsection{The physical block: Tricomi combination and its two operator towers}

The Ward identity has reduced the opposite-helicity light block to
Kummer's equation \eqref{eq:kummerode} for a single function
\(F(\chi)\). A second-order equation has two independent solutions,
so symmetry alone leaves one relative coefficient undetermined. In
this subsection we first write down the two solutions and show that
they are related by a reflection of the parameters of Kummer's
equation. We then show that the range of the
energy-fraction integral, which was fixed by the energy signs in the
previous subsection, imposes a growth condition at large \(\chi\) that
fixes the relative coefficient. The physical block turns out to be a
Tricomi function. Finally we reinstate the exchanged operator and
read off the two towers of light-basis operators that the block
contains. The two towers consist of light transforms of exchanged operators with scaling dimensions \(p+m\) and \(1+m\), respectively, where \(m=0,1,2,\ldots\). The second tower starts at the conformally soft value \(\Delta=1\).

\paragraph{The two Ward identity solutions.}
Near \(\chi=0\) we look for power-series solutions,
\(F=\chi^\varrho\sum_{m\geq0}c_m\chi^m\). Inserting this into
\eqref{eq:kummerode}, the lowest power of \(\chi\) fixes the exponent
\(\varrho\) and each higher power relates consecutive coefficients:
\begin{equation}
 \varrho(\varrho+p-1)=0,\qquad
 (m+\varrho+1)(m+\varrho+p)c_{m+1}=(m+\varrho+a)c_m.
 \label{eq:wardrecurrence}
\end{equation}
There are two possible leading powers, \(\varrho=0\) and \(\varrho=1-p\), and
for each of them the recurrence determines the whole series once
\(c_0\) is chosen. For generic \(p\) the two solutions are
\begin{equation}
 F_0={}_1F_1(a;p;\chi),\qquad
 F_1=\chi^{1-p}{}_1F_1(1-b;2-p;\chi),
 \label{eq:wardtwosolutions}
\end{equation}
where \({}_1F_1(a;c;\chi)=\sum_{m\geq0}\frac{(a)_m}{(c)_m\,m!}\chi^m\) is
the confluent hypergeometric function of the first kind and \((q)_m\) is
the rising factorial defined below \eqref{eq:completeblock}. The first solution is analytic at \(\chi=0\); the second
carries the extra power \(\chi^{1-p}\) and is not. Their Wronskian,
\((1-p)e^\chi\chi^{-p}\), is nonzero for noninteger \(p\), so they are
independent.

The second solution is not a new object. Once the power
\(\chi^{1-p}\) is factored out, what remains is again a Kummer
function, with the parameters of the first solution replaced by
\begin{equation}
 (a,p)\longmapsto(a-p+1,2-p)=(1-b,2-p).
 \label{eq:lightshadowreflection}
\end{equation}
The parameter \(p\) is replaced by \(2-p\), which has the
form of the shadow reflection of a dimension in two dimensions.
We therefore call \(F_1\) the \emph{light-shadow} solution.
Here the reflection relates the parameters of the two Kummer
functions. The external operators in the OPE remain fixed.

We work first with noninteger \(p\), for which the two displayed
solutions are independent. At integer \(p\), their separate
expressions can become singular. These values are obtained by
taking the limit of the full combination, which can produce
logarithmic terms.

\paragraph{The physical integration range.}
Symmetry has fixed the space of solutions, not the exact the
solution. What does is the range of the energy-fraction integral,
\(\tau\in(-\infty,0)\), which follows from positivity of the external
energies with \(\eps_1=-\eps_2\) and \(\eps_p=\eps_2\).
We now evaluate the Laplace transform
\(S=\int\dd \tau\,He^{-\varpi\tau}\) over this range and ask what it implies for
\(F\).

Work first in the chamber \(W/x<0\) and substitute \(\tau=(W/x)y\),
which maps \(\tau\in(-\infty,0)\) onto \(y\in(0,\infty)\) with positive
orientation and turns \(e^{-\varpi\tau}\) into \(e^{-\chi y}\). The
density then separates into the prefactor fixed above times a reduced
density \(w(y)\), for which \eqref{eq:lightdensityward} reads
\(\partial_y[y(1+y)w]-(a+py)w=0\), with solution
\(w=y^{a-1}(1+y)^{b-1}\) up to a constant absorbed into \(N_g\). Hence
\begin{equation}
 S(\varpi;W,x)=x^{-a}W^{p-1}F(\chi),\qquad
 F(\chi)=\int_0^\infty\dd y\,y^{a-1}(1+y)^{b-1}e^{-\chi y},
 \label{eq:Fhalfline}
\end{equation}
with the constant branch factor also absorbed into \(N_g\)
(Appendix~\ref{app:cycles}, eq.~\eqref{eq:orientedfusion}). The same
result follows from the explicit density \eqref{eq:rhosolution}
inserted in \eqref{eq:unknownlight}. \(F\) is an ordinary Laplace
integral on the positive half-line.

\textit{Growth condition.} Take first \(\operatorname{Re}a>0\) and
\(\operatorname{Re}\chi>0\), so that the integral converges at both
ends. For large positive \(\chi\) the exponential confines the
integrand to small \(y\), where the integrand behaves as
\(y^{a-1}\), and one finds \(F\sim\Gamma(a)\chi^{-a}\), where
\(\Gamma\) is Euler's gamma function. The physical
block is thus a power at large \(\chi\). In particular, it contains no
term growing like \(e^\chi\). We impose this growth condition for large positive \(\chi\).
Elsewhere, the block is defined by continuing the selected
solution together with its contour and branch prescription.

This growth condition plays the role here that single-valuedness
plays in Euclidean conformal field theory. It picks one combination
out of the two-dimensional solution space. To see how, write
\(F=C_0F_0+C_1F_1\). At large \(\chi\) both \(F_0\) and \(F_1\) grow \footnote{For generic parameters and large positive \(\chi\), \[ F_0\sim\frac{\Gamma(p)}{\Gamma(a)}e^\chi\chi^{-b}, \qquad F_1\sim\frac{\Gamma(2-p)}{\Gamma(1-b)}e^\chi\chi^{-b}. \]
These are the standard large-argument asymptotics and give precisely the coefficient written in \eqref{eq:growingmode}} like
\(e^\chi\) times the same power of \(\chi\), so their growing parts
can cancel against each other. Removing the common factor, the
coefficient of the growing exponential is
\begin{equation}
 C_0\frac{\Gamma(p)}{\Gamma(a)}+
 C_1\frac{\Gamma(2-p)}{\Gamma(1-b)}.
 \label{eq:growingmode}
\end{equation}
By requiring it to vanish, we can fix the ratio \(C_1/C_0\). On the other hand, requiring in addition
the normalization \(F\sim\Gamma(a)\chi^{-a}\) fixes both
coefficients:
\begin{equation}
 C_0=B(a,1-p),\qquad C_1=\Gamma(p-1),\qquad
 \frac{C_1}{C_0}=-\frac{\Gamma(p)\Gamma(1-b)}{\Gamma(a)\Gamma(2-p)}.
 \label{eq:shadowratio}
\end{equation}
Here \(B(x,y)=\Gamma(x)\Gamma(y)/\Gamma(x+y)\) is Euler's
beta function. The combination so
obtained is \(F=\Gamma(a)U(a,p,\chi)\), where \(U\) is Tricomi's
confluent hypergeometric function \cite{DLMF}. This is of course the
standard evaluation of the Laplace integral above. The point of the
derivation is that the two-term structure and the relative
coefficient were forced by symmetry and by the integration range.

Two remarks on normalization are in order. The coefficient of the leading
operator in the analytic solution is \(C_{12p}=N_gB(a,1-p)\). The
constant \(N_g\) includes the branch factor \(e^{\ip\phi_\Gamma}\), with
\(\phi_\Gamma=\pi(\varsigma_1-\varsigma_2)a\).
This factor carries the boundary prescriptions of the two external light kernels, as in \eqref{eq:cyclephase}.
In general it need not have unit modulus for complex \(a\).

The physical content of \eqref{eq:shadowratio} is the following.
For generic weights, the analytic solution \(F_0\) alone has an
exponentially growing contribution as \(\chi\to+\infty\).
The light-shadow solution cancels this contribution and gives
the power-law behaviour required by the convergent fusion
integral.

\paragraph{The operator block and its two towers.}
We now reinstate the exchanged operator. With the density and the
integration range fixed, the light block \eqref{eq:unknownlight}
becomes a definite operator-valued expression, which we denote by
\(\mathbb K_{12;p}^{L,U}\): the subscript names the external pair
and the exchanged family, and the superscript \(U\) reminds that the
Tricomi combination has been selected. Writing \(O_p\) in its
modified Mellin form \eqref{eq:mm}, the Laplace transform over the
energy fraction produces the Tricomi function inside the energy
integral, with argument \(\chi=\ip\eps_p\omega vW/x\):
\begin{align}
 \mathbb K_{12;p}^{L,U}
 ={}&N_gz_{12}^ax^{-a}\int_\Gamma\dd\bar w\,W^{p-1}
 \int_0^\infty\dd\omega\,\omega^{p-1}e^{-\ip\eps_p{\omega} u_2}
 \nn\\[-2pt]
 &\hspace{2cm}\times\Gamma(a)U(a,p,\ip\eps_p{\omega} vW/x)
 a_{-1}^{\eps_p}(\omega,z_2,\bar w).
 \label{eq:exactUblock}
\end{align}
The contour \(\Gamma\) is the light contour of leg~2, carried over
to the variable \(\bar w\). The energy integral runs over all
modes of the exchanged field, and the entire dependence on the time
separation \(v\) sits in the Tricomi function.

The full block in \eqref{eq:exactUblock} contains contributions
from both Ward identity solutions. We write these contributions separately as
\begin{equation*}
 \mathbb K_{12;p}^{L,U}
 =\mathbb K_{12;p}^{L,\mathrm{an}}
 +\mathbb K_{12;p}^{L,\mathrm{na}}.
\end{equation*}
Here \(\mathrm{an}\) denotes the contribution from \(F_0\),
which is analytic at \(\chi=0\), and \(\mathrm{na}\) denotes
the contribution from \(F_1\), which contains the factor
\(\chi^{1-p}\). Each contribution includes its coefficient
fixed in \eqref{eq:shadowratio}. Expanding the two solutions
gives two towers of light-basis operators. Each power \(\chi^m\) supplies
a factor \(\omega^m\), which raises the Mellin weight of the exchanged
operator from \(p\) to \(p+m\), and a factor \(W^m\), which combines
with \(W^{p-1}\) into the light kernel of weight \(p+m\). The
\(m\)-th term is therefore the light-transformed operator
\(\Aminus_{p+m}\) at the point of leg~2:
\begin{equation}
 \mathbb K_{12;p}^{L,\mathrm{an}}
 =C_{12p}z_{12}^ax^{-a}
 \sum_{m\geq0}\frac{(a)_m}{(p)_m m!}
 \left(\frac{\ip\eps_p v}{x}\right)^m\Aminus_{p+m}(2).
 \label{eq:completeblock}
\end{equation}
Here the argument \((2)\) denotes evaluation at \((u_2,z_2,\bz_2)\).
The rising factorial is defined by \((q)_0=1\) and
\((q)_m=q(q+1)\cdots(q+m-1)\) for \(m\geq1\).
These factors express the coefficients obtained from
\eqref{eq:wardrecurrence}. Before the light transform, a time
derivative raises the Mellin weight by one, as in \eqref{eq:mm}.
In \eqref{eq:completeblock}, each shift in Mellin weight is
accompanied by a change in the light kernel. The resulting
operators are therefore \(\Aminus_{p+m}\), with a different
light kernel at each level. 

The light-shadow solution gives the second tower:
\begin{equation}
 \mathbb K_{12;p}^{L,\mathrm{na}}
 =N_g\Gamma(p-1)z_{12}^ax^{-a}
 \left(\frac{\ip\eps_p v}{x}\right)^{1-p}
 \sum_{m\geq0}\frac{(1-b)_m}{(2-p)_m m!}
 \left(\frac{\ip\eps_p v}{x}\right)^m\Aminus_{1+m}(2).
 \label{eq:nonanalyticblock}
\end{equation}
The prefactor \(\chi^{1-p}\) lowers the weight of the leading term
from \(p\) to \(1\), so this tower starts at \(\Delta=1\), the
conformally soft dimension, and proceeds in integer steps from
there. It is distinguished from the first tower by the
non-integer power \(v^{1-p}\) of the time separation. Whether this
tower is subleading at small \(v\) depends on \(p\).
For generic \(p\) on the principal series,
\(\operatorname{Re}p=1\), the magnitude of \(v^{1-p}\)
is independent of \(|v|\) when the argument of \(v\) is held
fixed. The second tower therefore has no suppression by a
positive power of \(|v|\) relative to the first. Their relative
size also depends on their coefficients and operator matrix
elements. Its coefficient is fixed by the growth condition in \eqref{eq:shadowratio}.

Appendix~\ref{app:fusion} evaluates the fusion integral directly and
confirms both towers of the completed block.

\subsection{Same-helicity pair: the crossed block}
\label{sec:plusplus}

The four-gluon amplitude has a second crossed collinear channel, in
which the two operators that become collinear carry the same helicity.
We now repeat the construction of the previous two subsections for
this case. The steps are identical, with the following difference. Since both
external light transforms are now holomorphic, the angular dependence
of the exchanged operator survives the light transforms, and the
block depends on an angular frequency as well as on the time separation. This angular dependence must be retained when we later match the block to the same-helicity collinear channel and determine its large-spin contribution.

The relevant pair is \((3^+,1^+)\), with energy signs
\(\eps_3=+1\) and \(\eps_1=\eps_p=-1\).
For real, noncollinear momenta, the pair has spacelike total momentum. In analogy with \eqref{eq:abp} we set
\begin{equation}
 a_+=\Delta_3-1,\quad b_+=\Delta_1-1,\quad
 p_+=\Delta_3+\Delta_1-1=a_++b_++1,\quad
 v_+=u_3-u_1,\quad r_+=\bz_3-\bz_1,
 \label{eq:pluslabels}
\end{equation}
where \(v_+\) and \(r_+\) are the time and antiholomorphic
separations of the pair. The exchanged gluon has helicity \(+\) and
Mellin weight \(p_+\). For plus-helicity fields the barred weights
are \(2\bar h=\Delta-1\), so \(2\bar h_3=a_+\), \(2\bar h_1=b_+\) and
\(2\bar h_p=p_+-1\).The choice \(\eps_p=-1\) implies that the energies obey
\(\omega_p=\omega_1-\omega_3>0\).
The signed energy fraction carried by leg~3 is
\(\tau=-\omega_3/\omega_p\), so that
\(\omega_3=-\tau\omega_p\) and
\(\omega_1=(1-\tau)\omega_p\).
Positivity of the external energies therefore requires
\(\tau<0\).The Ward identity for \(\bar L_1\), computed exactly as
in \eqref{eq:parentwardaction}--\eqref{eq:rhoward} but with these
weights, gives the density equation and its solution
\begin{equation}
 \partial_\tau[\tau(1-\tau)\rho_+]
 -[a_+-(p_+-1)\tau]\rho_+=0,
 \qquad
 \rho_+=(-\tau)^{a_+-1}(1-\tau)^{b_+-1}.
 \label{eq:plusward}
\end{equation}
The parent kernel is therefore
\begin{equation}
 \mathbb K_{31;+}^{C}
 =N_g\delta(z_{31})\int_{-\infty}^{0}\dd\tau\,
 \rho_+(\tau)
 O_{p_+,+1}^{-}(u_1+\tau v_+,z_1,\bz_1+\tau r_+),
 \label{eq:plusparent}
\end{equation}
In $\mathbb K_{31;+}^{C}$, the subscript \(31\) identifies the external pair, and the \(+\) after the semicolon identifies the helicity of the exchanged gluon. The exchanged operator is inserted at the energy-weighted time and position.

We now apply the two light transforms. Both legs have positive
helicity, so both transforms integrate over the holomorphic
positions, with kernels \((w_3-z_3)^{a_+}\) and \((w_1-z_1)^{b_+}\).
The delta function \(\delta(w_3-w_1)\) sets the two integration
variables equal to a single \(w\), and the two kernels combine:
\begin{align}
 \mathbb K_{31;+}^{L}
 =N_g\int_\Gamma\dd w\,(w-z_3)^{a_+}(w-z_1)^{b_+}
 \int_{-\infty}^{0}\dd\tau\,\rho_+(\tau)
 O_{p_+,+1}^{-}(u_1+\tau v_+,w,\bz_1+\tau r_+).
 \label{eq:pluslightintegral}
\end{align}
Compare this with the opposite-helicity case. There the minus-leg
transform integrated over the antiholomorphic position, and
the exchanged operator's barred argument became the integration
coordinate \(\bar w\), with \(W=\bar w-\bz_2\). Here neither transform touches the barred coordinate, so the
exchanged operator still sits at \(\bz_1+\tau r_+\), and the energy
fraction \(\tau\) displaces it in time and in angle at the same time.
These shifts are generated by \(v_+\partial_{u_1}+r_+\partial_{\bz_1}\), acting on the exchanged field with \(v_+\) and \(r_+\) held fixed. To diagonalize this
operator we Fourier transform the exchanged field in its
antiholomorphic position, introducing the conjugate wave number
\(\xi\), the antiholomorphic counterpart of the variable of
Section~\ref{sec:conventions}, by
\(a_{+1}^{-}(\omega,w,\bz)=\int\dd\xi\,e^{\ip\xi\bz}
\widetilde a_{+1}^{-}(\omega,w,\xi)/(2\pi)\). On a mode of energy
\(\omega\) and angular frequency \(\xi\), the displacement
operator has the eigenvalue
\begin{equation}
 \chi_+=\ip\xi r_+-\ip\eps_p\omega v_+,
 \label{eq:pluschi}
\end{equation}
which plays the same role as \(\chi\) in the opposite-helicity case. Its time term has the sign fixed by the phase \(e^{-\ip\eps_p\omega(u_1+\tau v_+)}\). Each mode therefore enters the fusion integral with the factor \(e^{\tau\chi_+}\). With \(y=-\tau\) the energy-fraction integral becomes
\begin{equation}
 F_+(\chi_+)=\int_0^\infty\dd y\,y^{a_+-1}(1+y)^{b_+-1}e^{-\chi_+y}
 =\Gamma(a_+)U(a_+,p_+-1,\chi_+),\qquad
 \operatorname{Re}a_+>0,\quad\operatorname{Re}\chi_+>0,
 \label{eq:plusFhalfline}
\end{equation}
which, since \(a_++b_+=p_+-1\), satisfies \(\mathcal D_{a_+,p_+-1}F_+=0\), Kummer's equation with second parameter \(p_+-1\); the Tricomi function is again the combination selected by the growth condition of the previous subsection. Other weights and coordinates are reached by continuing this solution together with the integration contour and its boundary prescriptions. The completed block is
\begin{align}
 \mathbb K_{31;+}^{L,U}
 ={}&N_g\int_\Gamma\dd w\,(w-z_3)^{a_+}(w-z_1)^{b_+}
 \int_0^\infty\dd\omega\,\omega^{p_+-1}e^{-\ip\eps_p\omega u_1}
 \nn\\[-2pt]
 &\times\int\frac{\dd\xi}{2\pi}e^{\ip\xi\bz_1}
 \Gamma(a_+)U(a_+,p_+-1,\chi_+)
 \widetilde a_{+1}^{-}(\omega,w,\xi).
 \label{eq:plusUblock}
\end{align}
For noninteger \(p_+\), the Tricomi function separates into
the analytic solution and its light-shadow partner,
\begin{align}
 \Gamma(a_+)U(a_+,p_+-1,\chi_+)
 ={}&B(a_+,2-p_+){}_1F_1(a_+;p_+-1;\chi_+)\nn\\
 &+\Gamma(p_+-2)\chi_+^{2-p_+}
 {}_1F_1(1-b_+;3-p_+;\chi_+).
 \label{eq:plusconnection}
\end{align}
Since the Kummer parameter is now \(p_+-1\), the shadow exponent is \(2-p_+\) (in contract to \(1-p_+\)). Integer values of \(p_+\) are obtained by taking the limit of the full combination in \eqref{eq:plusconnection}.

Two points about this result will matter later. First, because
\(\chi_+\) contains \(\xi r_+\), the block is not a function
of the product \(\omega v_+\) alone. Expanding the analytic
solution in powers of \(\chi_+\) gives powers of
\(v_+\partial_{u_1}+r_+\partial_{\bz_1}\) acting on the
exchanged field. These contain time derivatives, angular
derivatives and products of both. The second solution also
retains the factor \(\chi_+^{2-p_+}\). A block depending
only on the time separation would lose this angular
dependence, which is needed to match the same-helicity
collinear channel in Section~\ref{sec:bridge}.

Second, the normalization is \(N_g=g/2\). The two light
kernels retain their boundary prescriptions inside the
\(w\)-integral. The factors associated with these prescriptions
therefore remain inside the integral, and no additional
constant factor is included in \(N_g\).

\section{The timelike two-particle spectrum}
\label{sec:partialwaves}
The blocks of the previous section describe what happens when two
operators with opposite energy signs become collinear. We now turn to
the other situation identified in Section~\ref{sec:conventions}, ie. a pair with equal energy signs and noncollinear momenta, whose total momentum is timelike. Such a pair does not fuse into a single massless gluon. Rather, it gives rise to massive two-particle states, labelled by their invariant mass and spin. This section constructs that two-particle basis for the physical scattering channel and fixes its normalization using the standard Jacob-Wick inner product. We then separate the tree amplitude into its two angular pole terms, which will be matched to the two crossed blocks in the next section, and finally carry the whole basis over to the finite-time light operators. Keeping the normalization of the spectrum independent of the crossed-channel matching is what will allow the matching in Section~\ref{sec:bridge} to be a crucial cross-check.

\subsection{Jacob-Wick expansion and the two pole sectors}

We work in the physical channel \(12\to34\), in which legs 1 and 2
have equal energy signs. Their total momentum has
\(M^2=s>0\), and the intermediate states are the two-gluon states
of invariant mass \(M\) and rest-frame spin \(J\). The Jacob-Wick
helicity functions resolve the two-gluon Fock space into these
massive Poincar\'e representations \cite{JacobWick,LawZlotnikov}, which is the two-particle sector announced in the introduction, and
we label it by \((\mu,J)\) with \(\mu=M^2\). For the sign pattern
\(\eps=(-,-,+,+)\) and all-outgoing helicities \((+,-,+,-)\), we
describe the kinematics at fixed \(s\) by the scattering angle,
\begin{equation}
 s=-(p_1+p_2)^2>0,\qquad t=-\frac{s}{2}(1-X),\qquad
 u_{\!M}=-\frac{s}{2}(1+X),\qquad X=\cos\theta,
 \label{eq:mandelstam}
\end{equation}
with \(t=-(p_1+p_3)^2\) and \(u_{\!M}=-(p_1+p_4)^2\).
Since the external particles are massless,
\(s+t+u_{\!M}=0\).

In the centre-of-mass frame, \(\theta\) is the angle between
the physical incoming momentum of gluon~1 and the outgoing
momentum of gluon~3. At \(\theta=0\), gluon~3 emerges along
the incoming direction of gluon~1. This is forward scattering,
with \(X=1\) and \(t=0\), where the pairs \((1,3)\) and
\((2,4)\) are collinear. At \(\theta=\pi\), gluon~3 emerges
in the opposite direction, along the incoming direction of
gluon~2. This is backward scattering, with \(X=-1\) and
\(u_{\!M}=0\), where the pairs \((1,4)\) and \((2,3)\) are
collinear. These two configurations lie at the ends of the
integration interval \(-1\leq X\leq1\). They give the
same-helicity and opposite-helicity collinear limits of
Section~\ref{sec:chiral}, respectively. In the initial and final
two-particle states the helicity differences are \(\lambda=-2\) and
\(\lambda'=2\), so only spins \(J\geq2\) occur. We fix the
one-particle normalization
\(\langle k,\ell|k',\ell'\rangle=2k^0(2\pi)^3
\delta^{(3)}(\mathbf k-\mathbf k')\delta_{\ell\ell'}\), for which
the two-body phase space measure is
\(\dd\Phi_2=\dd\Omega/(32\pi^2)\), with
\(\dd\Omega=\dd\varphi\,\dd X\), \(0\leq\varphi<2\pi\)
and \(-1\leq X\leq1\). Here \(\varphi\) is the azimuth of the
scattering plane. We use the Jacob-Wick conventions \cite{JacobWick}
for the Wigner rotation matrices, with the third Euler angle set to
zero,
\[
 D^J_{m\lambda}(\varphi,\theta,0)
 =e^{-\ip m\varphi}d^J_{m\lambda}(\theta),
\]
where \(m=-J,\ldots,J\) is the rest-frame spin projection and
\(d^J_{m\lambda}(\theta)\) is the matrix element for a rotation by
\(\theta\) about the \(y\)-axis. In the measure \(\dd\Omega\) the
functions \(\sqrt{8\pi(2J+1)}D^{J*}_{m\lambda}\) are orthonormal, the
star denoting complex conjugation.

Helicity amplitudes are functions of the scattering angle as well as a phase that depends on the little-group frames of the external spinors. The Jacob-Wick functions account for the angular dependence associated with the helicities, so what is expanded in them must be the amplitude with this spinor phase removed. For the present helicity configuration, the phase, and a convenient polynomial combination of spinor brackets, are
\begin{equation}
 \Phi=\frac{[13]\langle24\rangle}{\langle13\rangle[24]},
 \qquad \mathcal H=[13]^2\langle24\rangle^2=\Phi t^2,
 \label{eq:phase}
\end{equation}
and we expand \(A(s,X)=\cA/\Phi\). The expansion and its inverse,
with the normalizations just fixed, are
\begin{align}
 d_J(X)&=d^J_{-2,2}(\theta)
 =\frac{(1-X)^2}{4}P_{J-2}^{(4,0)}(X),\label{eq:dJ}\\
 \int_{-1}^1\dd X\,d_J(X)d_K(X)&=\frac{2\delta_{JK}}{2J+1},\nn\\
 A(s,X)&=16\pi\sum_{J\geq2}(2J+1)a_J(s)d_J(X),\nn\\
 a_J(s)&=\frac1{32\pi}\int_{-1}^1\dd X\,d_J(X)A(s,X).
 \label{eq:JW}
\end{align}

Here \(P_n^{(\alpha,\beta)}(X)\) is the Jacobi polynomial of
degree \(n\). For \(\alpha,\beta>-1\), these polynomials are
orthogonal on \([-1,1]\) with weight
\((1-X)^\alpha(1+X)^\beta\). We use the
normalization
\[
 P_n^{(\alpha,\beta)}(1)
 =\frac{\Gamma(n+\alpha+1)}{n!\,\Gamma(\alpha+1)}.
\]
In \eqref{eq:dJ}, \(n=J-2\), \(\alpha=4\) and \(\beta=0\).
In particular, \(P_{J-2}^{(4,0)}(1)=\binom{J+2}{4}\).

The projection integral applies directly when it converges. For the massless gluon amplitude below, the pole at \(X=-1\) makes this integral logarithmically divergent. Its partial-wave coefficients therefore require a subtraction prescription, which we specify in the next subsection.

Note the factor \((1-X)^2\) common to all basis functions. The initial and final helicity differences are unequal, so \(d^J_{-2,2}(0)=0\). For each fixed \(J\), \eqref{eq:dJ} gives a quadratic zero at \(X=1\). A finite sum retains this common factor, but an infinite sum need not vanish with the same power. We will return to this distinction at the end of the subsection.

The amplitude to be expanded is the tree four-gluon amplitude. Its
two colour-ordered components are the Parke--Taylor expressions
\cite{ParkeTaylor}
\begin{align}
 \cA_{1234}
 &=g^2\frac{\langle24\rangle^4}
 {\langle12\rangle\langle23\rangle\langle34\rangle\langle41\rangle}
 =-\frac{g^2\mathcal H}{s u_{\!M}},\nn\\
 \cA_{1324}
 &=g^2\frac{\langle24\rangle^3}
 {\langle13\rangle\langle32\rangle\langle41\rangle}
 =-\frac{g^2\mathcal H}{t u_{\!M}}.
 \label{eq:PT4}
\end{align}
Dividing by \(\Phi\) and using \eqref{eq:mandelstam} together with \(s+t+u_{\!M}=0\), the ordered amplitudes become rational functions of \(X\). After extracting the common factor \((1-X)^2\), the crossed-channel denominators are \(1+X\) and \(1-X\). Along with it, partial fractions separate the contributions associated with these two denominators,
\begin{equation}
 A_u(X)=\frac{g^2(1-X)^2}{2(1+X)},\qquad
 A_t(X)=-\frac{g^2(1-X)^2}{2(1-X)}=-\frac{g^2}{2}(1-X),
 \label{eq:polesectors}
\end{equation}
so that
\begin{align}
 A_{1234}&=A_u,\nn\\
 A_{1324}&=-\frac{g^2(1-X)^2}{2}
 \left[\frac1{1+X}+\frac1{1-X}\right]
 =-A_u+A_t=-g^2\frac{1-X}{1+X}.
 \label{eq:ordered}
\end{align}
Here, we name \(A_u\) and \(A_t\) the two \emph{pole sectors}. In
covariant form they are \(\cA_u=-g^2\mathcal H/(s u_{\!M})\) and
\(\cA_t=g^2\mathcal H/(s t)\). Each ordered amplitude is a signed
sum of pole sectors, and each pole sector contains exactly one of
the two crossed collinear singularities. The pole at \(s=0\) of \(\cA_{1234}\), associated with massless gluon exchange between the pairs \(12\) and \(34\), remains present in \(\cA_u\). At fixed \(s>0\), however, only the crossed invariants vary with the scattering angle. Also, the angular expansion resolves two-gluon states of positive invariant mass. Contributions supported only at \(\mu=0\), if present, are not determined by the spectrum on \(\mu>0\). They are treated differently as boundary data in Section~\ref{sec:inverse}. By the \emph{pole spectrum} of a sector
we mean the sequence of its partial-wave coefficients
\(a_J\). We will see in Section~\ref{sec:bridge} that the normalization of each pole spectrum is recovered from the corresponding crossed block. Finally, the colour-dressed amplitude is
\begin{equation}
 \bm\cA=c_s\cA_{1234}+c_t\cA_{1324},\qquad
 c_s=f^{a_1a_2b}f^{ba_3a_4},\quad
 c_t=f^{a_1a_3b}f^{ba_2a_4},
 \label{eq:DDM}
\end{equation}
Here \(f^{abc}\) are the gauge-group structure constants,
\(a_i\) is the adjoint colour index of leg \(i\), and the
repeated index \(b\) is summed. The tensors \(c_s\) and \(c_t\)
form the Del Duca-Dixon-Maltoni colour basis \cite{DDM}.

We can now explain a feature of \eqref{eq:polesectors} that will be important for the large-spin behaviour. The sector \(A_t\) is linear in \(X\). It is regular for \(-1\leq X\leq1\) and vanishes in the
forward limit \(X=1\). One might expect such a function to have only a few partial waves. However, it does not, because the basis functions \eqref{eq:dJ} all vanish to second order at \(X=1\), whereas \(A_t\) vanishes only to first order. The forward pole \(1/t\) of \(\cA_t\) cancels one of the two powers of \((1-X)\) supplied by the helicity tensor, not both. Dividing \(A_t\) by the common factor \((1-X)^2/4\) gives \(-2g^2/(1-X)\), which has a simple pole at \(X=1\). The function \(A_t\) itself remains regular for real scattering angles. The same-helicity collinear singularity of \(\cA_t\) is recovered after continuing \(z\) and \(\bz\) independently, as in Section~\ref{sec:plusplus}. A regular linear function thus requires infinitely many
helicity partial waves. Their large-spin behaviour is
determined by the simple pole that remains after dividing
by the common factor \((1-X)^2/4\).

\subsection{Normalized coefficients and collinear subtraction}

We now compute the partial-wave coefficients of the two pole
sectors. The forward sector \(A_t\) is a polynomial and can be
projected directly. The backward sector \(A_u\) cannot. This is because its pole sits at the endpoint \(X=-1\), where the basis functions \eqref{eq:dJ} do not vanish, so the projection integral diverges logarithmically. The backward sector therefore requires a subtraction prescription. We first displace the propagator pole, perform the projection, and subtract the logarithmic term before taking the massless limit.

Concretely, displace the two denominators to \(\zeta_u+X\)
and \(\zeta_t-X\), with \(\zeta_u,\zeta_t>1\), and write
\[
 A_u^{\zeta_u}(X)=\frac{g^2(1-X)^2}{2(\zeta_u+X)},
 \qquad
 A_t^{\zeta_t}(X)=-\frac{g^2(1-X)^2}{2(\zeta_t-X)}.
\]
At fixed \(s\), the displacement corresponds to a propagator
mass squared \(s(\zeta_r-1)/2\), where \(r=u,t\) labels the
pole sector. We denote the partial-wave coefficients of these
regulated amplitudes by
\[
 c_J^{r,\zeta_r}
 =\frac{1}{32\pi}\int_{-1}^{1}\dd X\,d_J(X)A_r^{\zeta_r}(X).
\]
To express these integrals, put \(n=J-2\) and define
\begin{equation}
\begin{aligned}
 \mathcal Q_n^{[4]}(\zeta)
 &=\frac1{32}\int_{-1}^{1}\dd y\,
   \frac{(1+y)^4P_n^{(0,4)}(y)}{\zeta-y},\\
 \mathcal P_n^{[4]}(\zeta)
 &=\frac1{32}\int_{-1}^{1}\dd y\,
   \frac{(1-y)^4P_n^{(4,0)}(y)}{\zeta-y},
 \qquad \zeta>1.
\end{aligned}
\label{eq:QPdefinitions}
\end{equation}
The superscript \([4]\) reminds of the fourth power of the angular factor in these integrals. \(\mathcal Q_n^{[4]}\)
and \(\mathcal P_n^{[4]}\) denote the integrated projections,
while \(P_n^{(\alpha,\beta)}\) denotes the Jacobi polynomial
inside each integral. Appendix~\ref{app:projections} derives their
closed forms and their limits as the regulators are removed. The projected coefficients are
\begin{equation}
 c_J^{u,\zeta_u}=\frac{g^2(-1)^J}{8\pi}
 \mathcal Q_{J-2}^{[4]}(\zeta_u),\qquad
 c_J^{t,\zeta_t}=-\frac{g^2}{8\pi}
 \mathcal P_{J-2}^{[4]}(\zeta_t),
 \label{eq:coeff}
\end{equation}
and by \eqref{eq:ordered} the coefficients of the two colour orderings are
\begin{equation}
 a_{J,1234}=c_J^{u},\qquad a_{J,1324}=-c_J^{u}+c_J^{t},
 \label{eq:orderedcoeff}
\end{equation}
with or without the regulator labels. When these labels are omitted,
\(c_J^u\) and \(c_J^t\) denote the massless sector coefficients with the subtraction prescription specified below. The sign \((-1)^J\) in the backward sector comes from the parity of the Jacobi polynomials under \(X\to-X\) and will be the signature of the backward collinear contribution throughout. Define
\[
 D_J=(J-1)J(J+1)(J+2),\qquad
 H_n=\sum_{m=1}^{n}\frac1m\quad(n\geq1),\qquad H_0=0.
\]
Here \(H_n\) is the \(n\)-th harmonic number. Writing
\(\delta=\zeta_u-1>0\), the limits as the regulators are removed at fixed \(J\):
\begin{equation}
 \mathcal Q_{J-2}^{[4]}(1+\delta)
 =\tfrac12[\log(2/\delta)-H_{J-2}-H_{J+2}]+o(1),
 \qquad \mathcal P_{J-2}^{[4]}(1)=\frac3{D_J}.
 \label{eq:Qlimits}
\end{equation}
The forward function has a finite limit, as anticipated. The
backward function diverges logarithmically, with a coefficient
independent of \(J\). This \(J\)-independence means that the
divergent part of the amplitude is supported at \(X=-1\). Indeed, writing \(\delta_{-1}=\delta(X+1)\), normalized by
\(\langle\delta_{-1},f\rangle=f(-1)\), and using
\(d_J(-1)=(-1)^J\), the divergent part of \(A_u\) is exactly the
distribution \(2g^2\log(2/\delta)\delta_{-1}\). Subtracting it
defines the finite part of the backward pole. This is the
standard plus-distribution extension
\(T_-(X)=[1/(1+X)]_+\), with
\(\langle T_-,f\rangle=\int_{-1}^1[f(X)-f(-1)]/(1+X)\dd X\). We denote the finite part after this subtraction by the
superscript \(\mathrm{FP}\). The resulting coefficients of
the two ordered amplitudes are
\begin{align}
 a_{J,1234}^{\rm FP}
 &=-\frac{g^2(-1)^J}{16\pi}(H_{J-2}+H_{J+2}),\nn\\
 a_{J,1324}^{\rm FP}
 &=\frac{g^2(-1)^J}{16\pi}(H_{J-2}+H_{J+2})
   -\frac{3g^2}{8\pi D_J}.
 \label{eq:finitecoeff}
\end{align}
These are the two sequences advertised in the introduction. Namely, the alternating harmonic-number sequence from the backward pole and the non-alternating rational sequence from the forward one.

The finite coefficients depend on how the backward pole is
defined as a distribution at \(X=-1\). This dependence is also present in
the spectral distribution in \((\mu,J)\). Changing the
subtraction prescription leaves the amplitude away from \(X=-1\) and its
collinear pole residue unchanged.

We parameterize the freedom in the finite subtraction by \(T_-^{[\kappa]}=T_-+\kappa\delta_{-1}\). We take \(\kappa\) to be independent of \(s\), so that the prescription preserves homogeneity in energy. Relative to \(\kappa=0\), this adds \(2g^2\kappa\delta_{-1}\) to \(A_{1234}\) and \(g^2\kappa(-1)^J/(16\pi)\) to \(a_{J,1234}\). The corresponding changes for the ordering \(1324\) have the opposite sign. The forward sector requires no subtraction because its projection converges. We return to the relation between the subtraction prescriptions in the two physical channels in Section~\ref{sec:endpointcrossing}.

\subsection{Timelike partial waves in the light basis}
\label{sec:inverse}

The partial-wave expansion above is a statement about momentum-space
amplitudes. To use it in the crossing equation we must express it in
the finite-time light basis, and this is the purpose of the present
subsection: we define the light-basis image of a partial wave of fixed invariant mass and spin, show that the full correlator is a superposition
of these images with the same coefficients as in momentum space, and
give the pairing that allows the coefficients to be read off.

Let \(\mu=M^2>0\). We define the timelike light block of invariant mass \(M\) and spin \(J\) by applying the transform \(\mathsf T\) of Section~\ref{sec:conventions} to the corresponding momentum-space partial wave. Including the normalization in \eqref{eq:JW} and the momentum-conservation delta function, we obtain
\begin{equation}
 \cB_{\mu J}^{L,s}
 =16\pi(2J+1)\mathsf T\!\left[
 (2\pi)^4\delta^{(4)}\!\left(\sum_{i=1}^4p_i\right)
 \delta(s-\mu)\Phi d_J(X)
 \right].
 \label{eq:massblock}
\end{equation}
The transformed block depends on all four Carroll times and celestial positions. Since \(\mathsf T\) is built from symmetry-covariant operations, these blocks transform under the global part of the BMS/ conformal Carroll algebra, exactly as the correlator does. The relevant Casimir is the usual quadratic Poincar\'e
Casimir, whose eigenvalue for a particle of mass \(m\) is
\(-m^2\) with our metric convention. For a two-particle
state, it is constructed from the total momentum.

Restoring the factors of \(\ip\) suppressed in
\eqref{eq:P}, the momentum generator on leg \(i\) is
\[
 \mathsf P_i^\nu=\ip q_i^\nu\partial_{u_i},
 \qquad q_i^\nu=q^\nu(z_i,\bz_i).
\]
Its action on the Fourier factor in \eqref{eq:mm} gives
the signed momentum \(p_i^\nu=\eps_i\omega_i q_i^\nu\).
Each external gluon is massless, so \(\mathsf P_i^2=0\).
However, the total momentum contains a cross term.
Using \(q_1\cdot q_2=-2z_{12}\bz_{12}\), we obtain
\[
 (\mathsf P_1+\mathsf P_2)^2
 =2\mathsf P_1\cdot\mathsf P_2
 =4z_{12}\bz_{12}\partial_{u_1}\partial_{u_2}
 \equiv4\cC_{12},
\]
which defines the pair operator \(\cC_{12}\), one quarter of the usual
Casimir for the pair. The factor \(\delta(s-\mu)\) in
\eqref{eq:massblock} selects total invariant mass squared
\(\mu\), so this Casimir has eigenvalue \(-\mu/4\).
Applying the light transforms gives
\begin{equation}
 \widehat{\cC}_{12}\cB_{\mu J}^{L,s}
 =-\frac\mu4\cB_{\mu J}^{L,s}.
 \label{eq:masscasimir}
\end{equation}
Here the hat denotes the action induced by the light
transforms, as in \eqref{eq:induced}. In the light
basis, as for every induced generator, the two time derivatives
raise the two weights by one and the celestial factors become
first-order operators. For the opposite-helicity light legs 1 and 2 the induced
Casimir is explicitly
\begin{equation}
 \widehat{\cC}_{12}{\cG^L}_{\boldsymbol\Delta}
 =-\eps_1\eps_2
 \left(1+\frac{z_{12}}{d_1}\partial_{z_1}\right)
 \left(\frac{\bz_{12}}{d_2}\partial_{\bz_2}-1\right)
 {\cG^L}_{\Delta_1+1,\Delta_2+1,\Delta_3,\Delta_4},\quad
 d_i=-\Delta_i.
 \label{eq:lightcasimir}
\end{equation}
Here \(d_i\) is the constant \(d=1-2h\) of \eqref{eq:lightP} for leg
\(i\).

To write the block explicitly we need to perform the energy integrals of \(\mathsf T\). For nondegenerate kinematics, momentum conservation fixes three energy ratios and imposes one condition on the angles, so the four Mellin integrals collapse to a single integral over an overall energy scale. Write
\(\omega_i=Y_iE\) for \(i=1,2,3\) and \(\omega_4=E\), with \(Y_i\) the
ratios fixed by momentum conservation, put \(s=\sigma E^2\), so that
\(\sigma\) is a purely angular quantity, and let
\(Q=\ip\eps_4(u_4+\sum_{i=1}^3(\eps_i/\eps_4)Y_i u_i)\) be the
combination of the four times that multiplies \(E\) in the Mellin
phase. Then, at fixed angular coordinates and for any function \(f\)
of the energies,
\begin{equation}
 \mathrm{MM}\!\left[(2\pi)^4\delta^{(4)}\!\Big(\sum_{i=1}^4p_i\Big)
 f(\omega_1,\omega_2,\omega_3,\omega_4)\right]
 =\int_0^\infty\dd E\,E^{\Lambda-1}e^{-QE}\,
 \mathcal J_\Delta\,f(Y_1E,Y_2E,Y_3E,E),
 \label{eq:Jdefinition}
\end{equation}
which defines the angular factor \(\mathcal J_\Delta\): it collects
the energy Jacobian, the Mellin factors
\(\prod_{i=1}^3Y_i^{\Delta_i-1}\), the positive-energy restrictions
and the remaining angular delta function, and is given explicitly in
\eqref{eq:jacobian}. Its subscript abbreviates the full set
\(\boldsymbol\Delta=(\Delta_1,\Delta_2,\Delta_3,\Delta_4)\). We first
work on the physical support, where \(Y_i>0\) and \(\sigma>0\). With
the time prescription \eqref{eq:timeboundary}, one has
\(\operatorname{Re}Q=\eta_4+\sum_{i=1}^3Y_i\eta_i>0\), which damps the
\(E\) integrand at large \(E\).

For the helicities \((+,-,+,-)\), the four light integrals
use the measure
\[
\begin{aligned}
 \dd\Gamma_L={}&
 \prod_{i\in\{1,3\}}
 \left[\dd w_i\,(w_i-z_i+\ip\varsigma_i0)^{\Delta_i-1}\right]\\
 &\times\prod_{j\in\{2,4\}}
 \left[\dd\bar w_j\,
 (\bar w_j-\bz_j+\ip\varsigma_j0)^{\Delta_j-1}\right].
\end{aligned}
\]
Each variable is integrated along its oriented light contour.
The angular quantities inside these integrals are evaluated at
\((Z_i,\bar Z_i)=(w_i,\bz_i)\) for \(i=1,3\), and at
\((Z_i,\bar Z_i)=(z_i,\bar w_i)\) for \(i=2,4\).
In particular, \(X\), \(\Phi\), \(Y_i\), \(\sigma\) and
\(\mathcal J_\Delta\) depend on these integration coordinates.
The delta function \(\delta(s-\mu)\) fixes
\(E=M/\sqrt\sigma\), and the block becomes
\begin{equation}
 \cB_{\mu J}^{L,s}
 =16\pi(2J+1)\int\dd\Gamma_L\,
 \mathcal J_\Delta\Phi d_J(X)
 \frac{M^{\Lambda-2}}{2\sigma^{\Lambda/2}}
 e^{-QM/\sqrt\sigma}.
 \label{eq:lightmasskernel}
\end{equation}
All the mass dependence, and all the dependence on the four Carroll
times, now sits in the explicit factor
\(M^{\Lambda-2}e^{-QM/\sqrt\sigma}\): at fixed light variables a
block of mass \(M\) is a single damped exponential in the times.

We now decompose the positive-mass contribution to the correlator. On \(s>0\), inserting \(1=\int_0^\infty\dd\mu\,\delta(s-\mu)\) into the momentum-space expansion and applying \(\mathsf T\) gives the first line below. The subtraction prescription is held fixed throughout. Acting with the pair Casimir then multiplies each block by \(-\mu/4\), as in \eqref{eq:masscasimir}. We denote the result by \(\cR^L=\widehat{\cC}_{12}\cG^L\) and call it the Casimir source.

A contribution with zero Casimir eigenvalue is annihilated. For example, its momentum-space kernel may be proportional to \(\delta(s)\), for which \(s\delta(s)=0\). A propagator pole \(1/(s+\ip0)\), however, obeys \(s/(s+\ip0)=1\) and contributes to the source. The expansion below describes the positive-mass part, with any terms supported only at \(\mu=0\) kept separately.
\begin{align}
\cG_\alpha^L&=\sum_{J\geq2}\int_0^\infty\dd\mu\,
 a_{J,\alpha}(\mu)\cB_{\mu J}^{L,s},\nn\\
 \cR_\alpha^L:=\widehat{\cC}_{12}\cG_\alpha^L
 &=-\frac14\sum_{J\geq2}\int_0^\infty\dd\mu\,
 \mu a_{J,\alpha}(\mu)\cB_{\mu J}^{L,s}.
 \label{eq:decomposition}
\end{align}
Here \(\alpha\) labels the colour ordering. As a check of the normalization, integrating the mass factor of \eqref{eq:lightmasskernel} over all masses must reproduce the energy integral of the correlator itself, and integrating it with the Casimir weight must reproduce the Casimir source. Both integrals are initially evaluated for \(\sigma>0\), \(\operatorname{Re}Q>0\) and \(\operatorname{Re}\Lambda>0\). The last condition ensures convergence at zero mass.
\begin{align}
 \int_0^\infty\dd\mu\,
 \frac{M^{\Lambda-2}}{2\sigma^{\Lambda/2}}e^{-QM/\sqrt\sigma}
 &=\Gamma(\Lambda)Q^{-\Lambda},\nn\\
 \int_0^\infty\dd\mu\left(-\frac\mu4\right)
 \frac{M^{\Lambda-2}}{2\sigma^{\Lambda/2}}e^{-QM/\sqrt\sigma}
 &=-\frac\sigma4\Gamma(\Lambda+2)Q^{-\Lambda-2}.
 \label{eq:radialcheck}
\end{align}
These expressions agree with the direct energy integrals for the correlator and its Casimir source. Other Mellin weights are reached by continuing the resulting expressions.

Within the regulated domain, Carroll-time derivatives of a fixed block can be taken under the integral whenever the differentiated integrals converge. Each \(\partial_{u_i}\) inserts a power of \(\omega_i\) and, by \eqref{eq:lightP}, is expressed through the block with the same \((\mu,J)\) and \(\Delta_i\) raised by one, acted on by a first-order operator in the celestial coordinate of leg \(i\). Outside this domain these derivatives are defined by continuation. Thus all time derivatives of the correlator are fixed by the same blocks labelled by \((\mu,J)\) and require no additional spectral coefficients.

To recover the momentum data, we restrict to the space of
physical momentum wave packets that can be reconstructed
uniquely from their complete light transforms. We call this
space the \emph{injective sector}.

Appendix~\ref{app:inverse} constructs this space using both boundary values of the light kernels introduced in Section~\ref{sec:conventions}. It contains finite sums over spins with smooth mass profiles of compact support inside \(\mu>0\). These packets provide the test functions for the distributional spectra considered below.

To define the pairing of two light images, we first undo \(\mathsf T\). We then remove the common momentum-conservation delta function and the helicity phase. Let's also denote the resulting momentum-space functions by \(f_1(s,X)\) and \(f_2(s,X)\). At fixed colour and physical helicity, we use the reduced measure \(\dd s\,\dd X\). The pairing and the resulting Gram matrix are
\begin{align}
 (f_1,f_2)_{\rm red}&=\int_0^\infty\dd s\int_{-1}^1\dd X\,f_1^*f_2,
 \label{eq:pairing}\\
 \mathsf G_{\mu J,\nu K}
 &=512\pi^2(2J+1)\delta_{JK}\delta(\mu-\nu),
 \label{eq:gram}
\end{align}
where the second line is the Gram matrix of the blocks \eqref{eq:massblock} in this pairing, diagonal in both labels, with
the measure \(\dd\mu\) and the normalization fixed by \eqref{eq:JW}.
If only one of the two boundary values of the light kernels is used,
\(\mathsf T\) acquires a kernel and the pairing is defined on the
quotient. This case is described in Appendix~\ref{app:inverse}.

With a diagonal Gram matrix the inversion is immediate. If the
Casimir source corresponds in momentum space to
\(\mathsf T^{-1}\cR^L=(2\pi)^4\delta^{(4)}(\sum p_i)\Phi \cR(s,X)\),
then, since \(\cR=-(s/4)A\),
\begin{equation}
 a_J(\mu)=-\frac1{8\pi\mu}\int_{-1}^1\dd X\,d_J(X)\cR(\mu,X).
 \label{eq:inversion}
\end{equation}
Three remarks on the scope of this formula. First, when the spectrum
is distributional rather than a smooth density, the formula is to be
read against smooth test functions compactly supported in \(\mu>0\),
and the angular integral uses the subtraction prescription of the
previous subsection. Second, contributions supported at
\(\mu=0\) are not recovered by it, because multiplication by \(\mu\)
can annihilate them. They are boundary data, to be supplied
separately. Third, the crossed massless blocks of
Section~\ref{sec:chiral} are not among the terms of the expansion:
they will be recovered from the positive-mass expansion by taking a
discontinuity in a crossed channel, which is the subject of the next
section.

Finally, Carroll-time homogeneity places a restriction on the
coefficients. From the explicit mass factor in
\eqref{eq:lightmasskernel}, each block satisfies
\begin{equation}
 \left(\sum_i u_i\partial_{u_i}+\Lambda-2
       -2\mu\partial_\mu\right)\cB_{\mu J}^{L,s}=0,
 \label{eq:blockscaling}
\end{equation}
that is, rescaling all times is equivalent to rescaling the mass. Suppose the correlator is homogeneous of degree
\(-(\Lambda+k)\) in the times, as in
\eqref{eq:homogeneity}. Applying the scaling operator to
\eqref{eq:decomposition}, using \eqref{eq:blockscaling},
and integrating by parts in \(\mu\) gives the coefficient
\(k a_J-2\mu\partial_\mu a_J\) inside the mass integral.
Invertibility of the transform therefore requires
\(2\mu\partial_\mu a_J=k a_J\) on \(\mu>0\).
For distributional coefficients, this relation is understood
against smooth test functions with compact support inside
\(\mu>0\). Compact support is imposed on these test
functions, so no boundary term from the mass integral enters this local relation.

For a homogeneous density, the boundary term in the full
mass integral vanishes initially when
\(\operatorname{Re}(\Lambda+k)>0\) and the time regulator
is retained. The resulting identity is then continued in
the Mellin weights. Appendix~\ref{app:massends} keeps the
boundary term explicit and explains the treatment at
Mellin poles.

For the Yang--Mills tree amplitude \(k=0\), so each
coefficient is constant in \(\mu\) at positive mass.
This fixes the energy dependence of the spectral data.
The uniqueness argument in Section~\ref{sec:spectralunique}
also uses crossing, the two crossed OPEs, the common subtraction prescription and the assumptions on the poles and growth
of the tree amplitude.
\section{Crossed OPEs, crossing and tree uniqueness}
\label{sec:crossing}

At this point we have two independently normalized objects. On one
side are the two crossed light blocks of Section~\ref{sec:chiral},
each describing a single massless gluon exchanged between a pair of
operators with opposite energy signs. On the other side is the
timelike spectrum of Section~\ref{sec:partialwaves}, the expansion
of the same four-point function in massive two-particle states of
the physical channel. This section relates them. We first show that
the two crossed blocks can be recovered from the timelike spectrum,
by summing the partial waves and taking the discontinuity at the
corresponding crossed pole. We call this relation between the two
expansions the \emph{bridge}. Since both pole sectors are determined
this way, both infinite sequences of spectral coefficients are
fixed by crossed OPE data, and we extract from them two exact
large-spin statements. We then write the crossing equation itself
and use it to test the freedom in the finite subtraction left over from Section~\ref{sec:partialwaves}. Finally we show that, for a fixed subtraction prescription, the two crossed OPEs, crossing and Carroll-time homogeneity determine the tree spectrum uniquely.

\subsection{Recovering both crossed light blocks from the spectrum}
\label{sec:bridge}

We begin by identifying which pairs of legs become collinear at the
forward and backward limits of the physical channel, and by making precise what it
means to extract an OPE from a correlator that is a distribution
rather than a function.

In the physical \(12\to34\) channel the backward limit
\(u_{\!M}=0\) is the collinear limit of the opposite-helicity pair
\((3^+,2^-)\), and the forward limit \(t=0\) is the collinear
limit of the same-helicity pair \((3^+,1^+)\). Both pairs have
opposite energy signs, so both are described by the crossed blocks
of Section~\ref{sec:chiral}. For the opposite-helicity pair the labels
\eqref{eq:abp} become
\begin{equation}
 a=\Delta_3-1,\quad b=\Delta_2,\quad
 p=\Delta_3+\Delta_2-1,\quad
 v=u_3-u_2,\quad x=\bz_2-\bz_3,\quad\eps_p=-1.
 \label{eq:crossedlabels}
\end{equation}
For the same-helicity pair the labels are those of
\eqref{eq:pluslabels}.

Before the light transforms, the parent kernels are supported on \(z_{3j}=0\) at fixed \(\bz_{3j}\), with \(j=2\) for the opposite-helicity pair and \(j=1\) for the same-helicity pair. To find such a term inside the four-point function we
cannot take a pointwise collinear limit. Instead we continue the
correlator away from the real locus \(\bz=z^*\), treating \(z\) and
\(\bz\) as independent variables, and take the discontinuity across
the holomorphic collinear pole. Define \(\operatorname{Disc}_zF\) as
the boundary value from above minus the boundary value from below.
Since \((z+\ip0)^{-1}-(z-\ip0)^{-1}=-2\pi\ip\delta(z)\), the
normalized discontinuities
\begin{equation}
 \mathscr D_u=-\frac1{2\pi\ip}\operatorname{Disc}_{z_{32}},
 \qquad
 \mathscr D_t=-\frac1{2\pi\ip}\operatorname{Disc}_{z_{31}},
 \qquad \frac{F}{z_{3j}}\longmapsto F\delta(z_{3j}),
 \label{eq:chiraldisk}
\end{equation}
turn a simple pole into a delta function with the same residue. Here \(F\) is regular in \(z_{3j}\) at zero, and its value there multiplies the delta function. All other coordinates and their boundary prescriptions are held fixed. Applied to the amplitude, the discontinuity produces the support of the parent kernels \eqref{eq:unknownparent} and \eqref{eq:plusparent}.

The comparison uses the energy sign \(\eps_p=-1\) of the exchanged operator chosen in Section~\ref{sec:chiral}. The wave packets are restricted to \(\omega_2>\omega_3\) for the opposite-helicity pair and to \(\omega_1>\omega_3\) for the same-helicity pair. On the injective sector, we define the corresponding operation on a light correlator by
\[
 \mathscr D_r^L=\mathsf T\,\mathscr D_r\,\mathsf T^{-1},
 \qquad r=u,t.
\]
Here \(\mathsf T^{-1}\) is the inverse of the complete transform on
the injective sector, constructed in Appendix~\ref{app:inverse}. 
The light contours retain the prescriptions of Appendix~\ref{app:cycles} at
every step.

The crossed block describes the exchanged operator, but a
four-point function also needs the coupling of that operator to the
two remaining legs. This coupling is the Carrollian three-point
kernel, the modified Mellin transform of the three-gluon amplitude,
\begin{equation}
 \cG_3^C(i,p,j)=\int_0^\infty
 \prod_{r\in\{i,p,j\}}\left[\dd\omega_r\,
 \omega_r^{\Delta_r-1}e^{-\ip\eps_r\omega_ru_r}\right]
 (2\pi)^4\delta^{(4)}(p_i+p_p+p_j)\cA_3(i,p,j).
 \label{eq:threepointkernel}
\end{equation}
The superscript \(C\) means that the modified Mellin transforms
have been performed, but the light transforms have not. The exchanged
operator has \(\Delta_p=p\) and helicity \(-1\) for the
opposite-helicity pair, or \(\Delta_p=p_+\) and helicity \(+1\) for
the same-helicity pair.
We use
\(\cA_3(i^+,j^-,k^-)=g\langle jk\rangle^3/
(\langle ij\rangle\langle ki\rangle)\), evaluated on the same
continued branch as the four-point kernel.

Pairing a block with this kernel means replacing its exchanged operator \(O_p(Y)\), inserted at the energy-weighted point \(Y\), by \(\cG_3^C\) with the exchanged leg at \(Y\). In the mode representation of the block the three-point momentum kernel replaces the mode, and in either form the Mellin integral of the exchanged leg is performed once. With \(\mathbb K_{32;p}^{L,U}\) obtained from \eqref{eq:exactUblock} by relabelling its external leg 1 as leg 3 and using \eqref{eq:crossedlabels}, the four-point contributions of the two complete exchanged families, including every descendant, are
\begin{equation}
 \mathcal F_u^L=\cL_1\cL_4\,
 \mathbb K_{32;p}^{L,U}\big[O_p(Y)\to\cG_3^C(1,p,4)\big],
 \qquad
 \mathcal F_t^L=\cL_2\cL_4\,
 \mathbb K_{31;+}^{L,U}\big[O_p(Y)\to\cG_3^C(p,2,4)\big].
 \label{eq:Fdefinition}
\end{equation}

We can now state the relation between the spectrum and the blocks. The operation \(\operatorname{Resum}_s^{\rm mer}\) first sums the displaced partial waves in momentum space and then takes \(\zeta_r\rightarrow1\) away from the collinear pole. The superscript \(\mathrm{mer}\) specifies the meromorphic part. Delta functions supported at collinear kinematics from the finite subtraction are not included in this operation. In the following we use the analytical continuation operation \(\operatorname{AC}\), which continues the resulting covariant kernel from the real slice \(\bz_i=z_i^*\) to independent \(z_i\) and \(\bz_i\). For a tree amplitude this continuation is explicit.
The resummed kernel is a rational function of the spinor brackets, which on the real slice are
\(\langle ij\rangle=\sqrt{2\eps_i\omega_i}\sqrt{2\eps_j\omega_j}\,
(z_j-z_i)\) and
\([ij]=\sqrt{2\eps_i\omega_i}\sqrt{2\eps_j\omega_j}\,
(\bz_j-\bz_i)\) up to the convention of Section~\ref{sec:conventions},
so \(\operatorname{AC}\) consists of substituting independent \(z_i\) and \(\bz_i\) into these brackets. The result is the unique analytic function of the eight variables that agrees with the kernel on the real slice. The momentum-conservation delta function is treated as in Appendix~\ref{app:cycles}. It is imposed first on its real support, and the integration region it selects is then continued together with the light contours.

When these operations act on a light correlator, they are understood through \(\mathsf T\). In particular, the left side of the bridge formula below uses the convention
\[
\begin{aligned}
 \mathscr D_r^L\operatorname{AC}\operatorname{Resum}_s^{\rm mer}
 [\cG^L]
 =
 \mathsf T\!\left[
 \mathscr D_r\operatorname{AC}\operatorname{Resum}_s^{\rm mer}
 [\mathsf T^{-1}\cG^L]\right].
\end{aligned}
\]
The complete transform is restored after the discontinuity is taken.

\begin{theorem}[Both crossed light blocks from the timelike spectrum]
\label{thm:bridge}
After resummation, the timelike partial waves reproduce the
complete crossed light block at each pole.
For the Yang--Mills cubic OPE data, the two pole spectra
\eqref{eq:coeff} and the timelike light blocks \eqref{eq:massblock}
satisfy
\begin{equation}
 \boxed{\displaystyle
 \mathscr D_r^L\operatorname{AC}\operatorname{Resum}_s^{\rm mer}
 \left[\sum_{J\geq2}\int_0^\infty\dd\mu\,
 c_J^{r,\zeta_r}\cB_{\mu J}^{L,s}\right]
 =\mathcal F_r^L,\qquad r=u,t.}
 \label{eq:bridgetheorem}
\end{equation}
The identity holds componentwise for the cycles of Appendix~\ref{app:cycles}, first in their common convergence domain and then by the ordered operations just specified. The right sides are the four-point contributions \eqref{eq:Fdefinition} formed from the complete crossed light blocks \eqref{eq:exactUblock} and \eqref{eq:plusUblock}. All descendants, in Carroll time and in the untransformed angular coordinate, are retained. With a common subtraction prescription, the ordered coefficients follow from  \eqref{eq:orderedcoeff} and they are the
coefficients \(a_{J,\alpha}\) that enter the spectral expansion
\eqref{eq:decomposition} of the light correlator.
\end{theorem}

The argument has two halves. The first half shows that the partial
waves resum to the pole sectors. The second half shows that the
discontinuity of each pole sector is the corresponding parent kernel.

For the first half we need the sums of the displaced-pole
coefficients against the basis functions. They are
\begin{align}
 2\sum_{J\geq2}(2J+1)(-1)^J\mathcal Q_{J-2}^{[4]}(\zeta)d_J(X)
 &=\frac{(1-X)^2}{2(\zeta+X)},\nn\\
 2\sum_{J\geq2}(2J+1)\mathcal P_{J-2}^{[4]}(\zeta)d_J(X)
 &=\frac{(1-X)^2}{2(\zeta-X)},
 \label{eq:heinebridge}
\end{align}
with the coefficients taken from the normalized projections of Appendix~\ref{app:projections}. Multiplying the first identity by \(g^2\) and the second by \(-g^2\) gives the two displaced pole sectors of \eqref{eq:polesectors}. Inserting them into the spectral decomposition and performing the mass integral with \eqref{eq:radialcheck} gives the light transform of the two covariant pole sectors. In the common starting domain of the contours, the displaced sums converge uniformly on compact physical angular intervals and the mass integral converges absolutely, so this equality holds on matrix elements between wave packets before any continuation. The continuation to the crossed discontinuity and to the finite-time boundary values then follows from the transported prescriptions of Appendix~\ref{app:cycles}.

For the second half we use collinear factorization. Consider first
the opposite-helicity pair and parametrize the collinear spinors as
\(\lambda_3=\sqrt\tau\lambda_p\) and
\(\lambda_2=\sqrt{1-\tau}\lambda_p\). The universal collinear limit
of the four-gluon amplitude \cite{BDDK} is
\begin{equation}
 \cA_4(1^+,2^-,3^+,4^-)
 \sim g\frac{(1-\tau)^{3/2}}{\sqrt\tau\langle23\rangle}
 \cA_3(1^+,p^-,4^-).
 \label{eq:splitbridge}
\end{equation}
On the energy branch selected in Section~\ref{sec:chiral} one has
\(\omega_3=-\tau\omega_p\) and \(\omega_2=(1-\tau)\omega_p\) with
\(-\infty<\tau<0\), so that \(\tau\) is the signed energy fraction
of leg 3. The collinear bracket is
\(\langle23\rangle=2\eps_p\omega_p\sqrt{\tau(1-\tau)}z_{32}\), and
the change of variables from \((\omega_2,\omega_3)\) to
\((\omega_p,\tau)\) has Jacobian \(\omega_p\). Collecting the
Mellin weights of legs 2 and 3, the measure of the collinear term
becomes
\begin{equation}
 \frac{g}{2z_{32}}\dd\omega_p\,\omega_p^{p-1}
 \dd\tau\,(-\tau)^{a-1}(1-\tau)^b
 e^{-\ip\eps_p\omega_p(u_2+\tau v)}.
 \label{eq:measurebridge}
\end{equation}
The energy-fraction density in this expression is \(\rho(\tau)=(-\tau)^{a-1}(1-\tau)^b\), which solves \eqref{eq:rhoward}. It is the density of the parent kernel, while the holomorphic factor is still \(1/z_{32}\). Applying \(\mathscr D_u\) replaces this pole by \(\delta(z_{32})\) and gives \eqref{eq:unknownparent}, with the relabelling in \eqref{eq:crossedlabels}. The comparison fixes \(N_g^{\rm parent}=g/2\). After the two light transforms, the constant in \eqref{eq:exactUblock} is \(N_g^\Gamma=(g/2)e^{\ip\phi_\Gamma}\), with the contour factor specified in \eqref{eq:cyclephase}. This proves the case \(r=u\).

For the same-helicity pair we parametrize instead
\(\lambda_3=\sqrt\tau\lambda_p\) and
\(\lambda_1=\sqrt{1-\tau}\lambda_p\). Legs 1 and 3 are adjacent in
the \(1324\) ordering, whose collinear limit is
\begin{equation}
 \cA_4(1^+,3^+,2^-,4^-)
 \sim\frac{g}{\sqrt{\tau(1-\tau)}\langle13\rangle}
 \cA_3(p^+,2^-,4^-).
 \label{eq:plussplitting}
\end{equation}
This follows directly from \eqref{eq:PT4} and the three-point
amplitude in \eqref{eq:threepointkernel}. Now
\(\omega_3=-\tau\omega_p\), \(\omega_1=(1-\tau)\omega_p\), and
\(\langle13\rangle=2\eps_p\omega_p\sqrt{\tau(1-\tau)}z_{31}\), so
the splitting factor equals
\(g/[2\omega_p(-\tau)(1-\tau)z_{31}]\). Multiplying by the energy
Jacobian and the Mellin weights of legs 1 and 3 gives
\begin{equation}
 \frac{g}{2z_{31}}\dd\omega_p\,\omega_p^{p_+-1}
 \dd\tau\,(-\tau)^{a_+-1}(1-\tau)^{b_+-1}
 e^{-\ip\eps_p\omega_p(u_1+\tau v_+)}.
 \label{eq:plusmeasure}
\end{equation}
The density is the one found in \eqref{eq:plusward}, and momentum
conservation places the exchanged operator at
\(\bz_p=\bz_1+\tau r_+\), as in \eqref{eq:plusparent}. Applying
\(\mathscr D_t\) therefore gives the parent kernel
\eqref{eq:plusparent} with \(N_g=g/2\). The two holomorphic light
integrals then give \eqref{eq:pluslightintegral}, and the half-line
integral over the energy fraction gives
\(\Gamma(a_+)U(a_+,p_+-1,\chi_+)\), which is \eqref{eq:plusUblock}.
One point deserves attention here. The \(1234\) ordering has no
\(\langle13\rangle\) pole, so it is regular in this collinear limit
and the entire residue comes from the sector \(\cA_t\). This proves
the case \(r=t\).

The theorem has a simple takeaway. In the crossed channel the exchanged operator is massless. The crossed Casimir of the parent kernel vanishes because its delta function is supported where the crossed invariant is zero. In the physical channel the same contribution is spread over the timelike continuum.

At fixed nonzero \(s\), a finite sum of the covariant partial waves \(\Phi d_J(X)\) has no crossed collinear pole and therefore has zero chiral residue. The crossed block is recovered after the infinite sum is performed and the covariant result is continued. In the same-helicity sector, the stripped function \(A_t(X)\) is linear. The collinear pole belongs to \(\cA_t=\Phi A_t=g^2\mathcal H/(s t)\), which is the quantity continued to independent spinor coordinates.

\subsection{Exact large-spin tails}

Theorem~\ref{thm:bridge} says that each pole sector, and therefore
each of the two infinite sequences in \eqref{eq:finitecoeff}, is
fixed by one crossed OPE. We now make this explicit at large spin,
where the two sequences can be compared with the asymptotic data a
bootstrap would use.

The backward pole produces the entire harmonic-number sequence in
\eqref{eq:finitecoeff}. Its expansion at large spin is
\begin{equation}
 H_{J-2}+H_{J+2}
 =2\log J+2\gamma_E+\frac1J-\frac{25}{6J^2}+O(J^{-3}),
 \label{eq:largeJbridge}
\end{equation}
where \(\gamma_E=\lim_{n\to\infty}(H_n-\log n)\) is Euler's constant. Not only the logarithm but every inverse power of \(J\) in this expansion is determined by the same crossed OPE, because the sequence comes from a single pole. The spectral coefficients also contain the subtraction constant \(\kappa\) of Section~\ref{sec:partialwaves}. It shifts \(a_{J,1234}\) by \(g^2\kappa(-1)^J/(16\pi)\), without changing the harmonic-number expansion above. Remove the alternating sign by defining \(\widehat a_J=(-1)^J[a_{J,1234}^{\rm FP} +g^2\kappa(-1)^J/(16\pi)]\). The \(\kappa\) contribution to \(\widehat a_J\) is independent of \(J\), so it cancels in differences of neighbouring spins. The result is exact,
\begin{equation}
 \boxed{\displaystyle
 \widehat a_{J+1}-\widehat a_J
 =-\frac{g^2}{16\pi}
 \left(\frac1{J-1}+\frac1{J+3}\right),\qquad J\geq2.}
 \label{eq:adjacentspin}
\end{equation}
This relation is independent of the subtraction prescription, and it
measures the residue of the crossed pole through the spectral
coefficients at every pair of adjacent spins.

The forward contribution has a convergent angular projection,
which can be evaluated directly. One finds
\begin{equation}
 c_J^t=-\frac{g^2}{64\pi}\int_{-1}^1\dd X\,d_J(X)(1-X)
 =-\frac{3g^2}{8\pi D_J}.
 \label{eq:forwardtail}
\end{equation}
The sum of these coefficients against the basis reproduces the
forward sector, and their expansion at large spin is a pure power
series,
\begin{align}
 \sum_{J\geq2}\frac{(2J+1)d_J(X)}{D_J}&=\frac{1-X}{12},\nn\\
 c_J^t&=-\frac{3g^2}{8\pi J^4}
 \left(1-\frac2J+\frac5{J^2}-\frac{10}{J^3}+O(J^{-4})\right).
 \label{eq:forwardresum}
\end{align}
This is the complete non-alternating part of \(a_{J,1324}\). Its overall coefficient, its sign, and every subleading power are fixed by the same-helicity OPE. The two infinite sequences in \eqref{eq:finitecoeff} therefore have different origins, corresponding to one operator in each crossed block. Combining them with the signs in \eqref{eq:ordered} gives the spectrum of each colour-ordered amplitude.

\subsection{Crossing and the common subtraction constant}
\label{sec:endpointcrossing}

The bridge theorem relates the spectrum of one channel to the
crossed blocks. Crossing symmetry relates the spectra of two
channels to each other. We now write that relation in the light
basis and use it for one concrete purpose, to determine whether it fixes the subtraction constant \(\kappa\).

The two physical channels we compare are \(12\to34\) and
\(13\to24\). In the original leg labels their energy signs are
\((-,-,+,+)\) and \((-,+,-,+)\), respectively. Each channel has its
own helicity phase, removed before projecting: if \(i,j\) are the
plus-helicity legs and \(k,l\) the minus-helicity legs, the phase is
\([ij]\langle kl\rangle/(\langle ij\rangle[kl])\), evaluated in that
channel's leg ordering. We write \(\Phi_s\) and \(\Phi_t\) for the
two phases, so that \(\Phi_s\) is the phase \(\Phi\) of
\eqref{eq:phase}.

The channel \(13\to24\) needs its own angular variables,
\begin{equation}
 t=\nu>0,\qquad s=-\frac\nu2(1-X_t),\qquad
 u_{\!M}=-\frac\nu2(1+X_t),
 \label{eq:tchannelangles}
\end{equation}
where \(\nu\) is the invariant mass squared of this channel and
\(X_t=1+2s/t\) the cosine of its scattering angle. The pairs
\((1^+,3^+)\) and \((2^-,4^-)\) have helicity differences \((0,0)\),
so the partial-wave basis consists of the Legendre polynomials
\(P_K(X_t)=P_K^{(0,0)}(X_t)\), normalized by \(P_K(1)=1\), with
\(K\geq0\). The timelike light block \(\cB^{L,t}_{\nu K}\) of this
channel is defined as in \eqref{eq:massblock}, with \(\delta(t-\nu)\),
the phase \(\Phi_t\), the angular function \(P_K(X_t)\) and the energy
signs of the \(13\to24\) channel. We write
\(a^s_{J,\alpha}\equiv a_{J,\alpha}\) for the coefficients of
Section~\ref{sec:partialwaves}, with \(J\geq2\), and \(a^t_{K,\beta}\)
for those of the crossed channel, where \(\alpha,\beta\) label the
helicity and colour components.

We will now derive the crossing equation involving the crossing operator \(\mathcal X_{st}\). Four pieces of data are required to fix it, as listed below. Let \(\pi_{23}\) be
the permutation that exchanges legs 2 and 3, including their momenta,
helicities and colour indices. First, \(\mathcal X_{st}\) contains
this permutation. Second, it contains the analytic continuation from
the physical region of the \(12\to34\) channel to that of the
\(13\to24\) channel, along the boundary prescription of
Appendix~\ref{app:cycles}. For a tree amplitude this continuation is
the explicit substitution described after \eqref{eq:Fdefinition}.
Third, on helicity-stripped amplitudes it contains the ratio of the
two helicity phases. Fourth, on the colour basis \((c_s,c_t)\) of
\eqref{eq:DDM} the permutation exchanges the two tensors. In
formulas,
\begin{equation}
 \mathcal X_{st}A=\frac{\Phi_t\circ\pi_{23}}{\Phi_s}\,(A\circ\pi_{23}),
 \qquad
 \mathcal X_{st}\colon
 \begin{pmatrix}c_s\\c_t\end{pmatrix}\longmapsto
 \begin{pmatrix}0&1\\1&0\end{pmatrix}
 \begin{pmatrix}c_s\\c_t\end{pmatrix},
 \label{eq:crossingoperator}
\end{equation}
the continuation being understood. On a light correlator the same
operator is defined through the transform, in the same way as the
discontinuity \(\mathscr D_r^L\) above,
\begin{equation}
 \mathcal X_{st}\,\cG^L
 =\mathsf T\!\left[\operatorname{AC}\!\left(
 \frac{\Phi_t\circ\pi_{23}}{\Phi_s}\,
 (\mathsf T^{-1}\cG^L)\circ\pi_{23}\right)\right].
 \label{eq:crossinglight}   
\end{equation}
Three properties follow from this definition and can be used as
checks. Applied twice, \(\mathcal X_{st}\) returns the original
correlator, since \(\pi_{23}\) is an involution and the two
continuations retrace each other. It commutes with the induced
generators of Section~\ref{sec:conventions}, because the permutation
and the continuation are both Poincar\'e covariant, and it therefore
maps blocks covariant under Ward identity of one channel to that
of the other. And it conjugates the pair Casimir of the \(12\)
channel into the pair Casimir of the \(13\) channel, so that it maps
a block of definite \(\mu\) in one channel to a superposition of
blocks of all \(\nu\) in the other. The light blocks are built from
the covariant amplitudes, so the helicity phases are already inside
them. The crossing equation in the light basis then reads
\begin{equation}
 \boxed{\displaystyle
 \sum_{J,\alpha}\int_0^\infty\dd\mu\,
 a^s_{J,\alpha}(\mu)\cB^{L,s}_{\mu J,\alpha}
 =\mathcal X_{st}\!
 \left[\sum_{K,\beta}\int_0^\infty\dd\nu\,
 a^t_{K,\beta}(\nu)\cB^{L,t}_{\nu K,\beta}\right].}
 \label{eq:crossing}
\end{equation}
Because both sides are expanded in normalized blocks with the
pairing of Section~\ref{sec:inverse}, this is an equation between
definite spectral coefficients rather than between unnormalized
functions. On the injective sector it is equivalent, after undoing
the transform, to ordinary helicity crossing of the amplitude. What the light basis adds is on the input side. In this basis, the collinear
factorization data \footnote{By the collinear factorization data of the amplitude we mean the residues at its two
collinear poles together with the full dependence of the exchanged
gluon on the kinematics of the pair.} becomes explicitly usable.  These data are the momentum-space
counterpart of the operator-product data at null infinity. The
crossed blocks provide them in the two collinear limits, with their
complete dependence on the time and angular separations, which a pole residue alone does not specify.

We now use \eqref{eq:crossing} to test the subtraction constant. Fix
the same \(\kappa\) per unit angular pole residue in every sector.
In the \(s\) channel the two ordered amplitudes are then
\begin{align}
 A_{1234}^{[\kappa]}&=\frac{g^2}{2}(1-X)^2T_-(X)
 +2g^2\kappa\delta_{-1},\nn\\
 A_{1324}^{[\kappa]}&=-g^2(1-X)T_-(X)
 -2g^2\kappa\delta_{-1}.
 \label{eq:s_endpointfamily}
\end{align}
These distributions obey \(A_{1234}^{[\kappa]}+(1-X)A_{1324}^{[\kappa]}/2=0\), which extends the ordering relation \eqref{eq:ordered} to \(X=-1\). If separate subtraction constants are assigned to the two orderings, requiring this relation to remain valid makes them equal. Thus the common choice is compatible with the ordering identity as well as with the crossing test below.

In the angular variables \eqref{eq:tchannelangles} of the crossed
channel \(13\to24\), the helicity-stripped ordered amplitudes are
\begin{equation}
 A^t_{1234}=-2g^2\left(\frac1{1-X_t}+\frac1{1+X_t}\right),
 \qquad A^t_{1324}=\frac{2g^2}{1+X_t}.
 \label{eq:tchannelamplitudes}
\end{equation}
We denote the coefficients in this channel by
\(b_{J,\alpha}=a^t_{J,\alpha}\), with \(\alpha\) the colour
ordering, normalized as in \eqref{eq:JW},
\[
 A^t_\alpha(X_t)
 =16\pi\sum_{J\geq0}(2J+1)b_{J,\alpha}P_J(X_t),\qquad
 b_{J,\alpha}
 =\frac1{32\pi}\int_{-1}^{1}\dd X_t\,P_J(X_t)A^t_\alpha(X_t).
\]
The projection integrals are understood with subtractions at the collinear poles. For a smooth test function \(f\), the forward prescription is defined by
\[
 \int_{-1}^{1}\dd X_t\,f(X_t)
 \left[\frac1{1-X_t}\right]_+
 =\int_{-1}^{1}\dd X_t\,\frac{f(X_t)-f(1)}{1-X_t}.
\]
The backward prescription subtracts \(f(-1)\) instead, as in the definition of \(T_-\). At tree level these coefficients do not depend on \(\nu\), by the same homogeneity argument as in Section~\ref{sec:inverse}. Both the forward and backward limits of this channel contain massless poles, and we extend both with the same prescription and the same constant \(\kappa\). This gives
\begin{align}
 b_{J,1234}&=\frac{g^2}{8\pi}(1+(-1)^J)H_J
 -\frac{g^2\kappa}{16\pi}(1+(-1)^J),\nn\\
 b_{J,1324}&=-\frac{g^2(-1)^J}{8\pi}H_J
 +\frac{g^2\kappa(-1)^J}{16\pi}.
 \label{eq:crossedfinitecoeff}
\end{align}
The two projection identities are
\[
\begin{aligned}
 \int_{-1}^{1}\dd X_t\,P_J(X_t)
 \left[\frac1{1-X_t}\right]_+&=-2H_J,\\
 \int_{-1}^{1}\dd X_t\,P_J(X_t)
 \left[\frac1{1+X_t}\right]_+&=-2(-1)^JH_J.
\end{aligned}
\] For a displaced pole the projection is the Legendre function of the second kind,
\[
 Q_J(\zeta)=\frac12\int_{-1}^{1}\dd X_t\,
 \frac{P_J(X_t)}{\zeta-X_t},\qquad \zeta>1,
\] and the limit \(\zeta\to1^+\) is taken afterwards, as in Ref.~\cite{LawZlotnikov}.

These formulas give a family of subtraction prescriptions compatible with crossing and the ordering identity for every common value of \(\kappa\). Within the prescription used here, the same constant is assigned per unit pole residue. The calculation does not classify all possible distributions supported at \(X_t=\pm1\).

To write this common freedom covariantly, express a crossed pole as \(\bm N_q/q\). Here \(q\) is one of \(-s,-t,-u_{\!M}\), chosen positive on the crossed physical side of the pole. For example, \(q_u=-u_{\!M}=s(1+X)/2\). The numerator \(\bm N_q\) carries the colour indices and is regular in \(q\) near zero. Its value \(\bm N_q|_{q=0}\) is the residue, with the other independent kinematic variables held fixed. A change from the reference choice \(\kappa=0\) changes the amplitude by
\begin{equation}
 \delta\bm\cA=\kappa\sum_{q\,\mathrm{channels}}
 \bm N_q\big|_{q=0}\delta(q).
 \label{eq:covariantendpoint}
\end{equation}
This expression is crossing covariant and has energy degree zero. Under a common energy rescaling, \(\bm N_q\) has degree two and \(\delta(q)\) has degree minus two. The added term is supported on the collinear boundary. It changes the distribution at the pole and its spectral coefficients, while leaving unchanged the meromorphic residues extracted in \eqref{eq:bridgetheorem}. Thus the common value of \(\kappa\) remains part of the subtraction prescription.

\subsection{Tree uniqueness from both crossed OPEs}
\label{sec:spectralunique}
The results so far can be read as a sequence of statements about the
crossing equation \eqref{eq:crossing}. The bridge theorem
\eqref{eq:bridgetheorem} shows that the two collinear limits of the
amplitude, \eqref{eq:splitbridge} and \eqref{eq:plussplitting}, fix
the two infinite sequences \eqref{eq:finitecoeff} of spectral
coefficients in the physical channel. The test of
Section~\ref{sec:endpointcrossing} shows that the Yang--Mills
coefficients \eqref{eq:finitecoeff} and \eqref{eq:crossedfinitecoeff}
satisfy \eqref{eq:crossing}. The question we now ask is whether they
are the only solution. More precisely, we fix the inputs that a
bootstrap would be given, the two crossed operator products with
their normalization, the subtraction prescription of
\eqref{eq:s_endpointfamily}, and the time scaling of the correlator,
and we ask how many pairs of spectra \((a^s_J,a^t_K)\) satisfy
\eqref{eq:crossing} with these inputs. If the answer is one, the
crossing equation determines the tree amplitude from its collinear
data.
 
The question must be asked within a class of amplitudes, and the
natural class is the one to which tree amplitudes belong. After the
regular helicity tensor \(\mathcal H\) of \eqref{eq:phase} is
extracted, the Parke--Taylor amplitudes \eqref{eq:PT4} are rational
functions of the Mandelstam invariants whose only singularities are
simple poles on the collinear loci. We therefore define the rational
tree class as the set of amplitudes of the form
\begin{equation}
 \cA=\mathcal H\,R(s,t,u_{\!M}),\qquad
 R\ \text{rational, with simple poles only at }s,t,u_{\!M}=0,\qquad
 |R|\leq C\,(1+|s|+|t|)^N
 \label{eq:treeclass}   
\end{equation}
on complex kinematics, for some constant \(C\) and integer \(N\).
Locality is what restricts the poles to the collinear loci, and the
polynomial bound is the tree-level statement that no further
singularities are hidden at infinity. Within this class the residues
of all poles are fixed by the two crossed collinear types and their
permutations, so two members with the same crossed operator products
differ by an amplitude without poles.
 
Consider then two members of the class with the same crossed
operator products and the same subtraction prescription, and take
their difference \(\delta\cA\). It has no poles, so after the
helicity tensor is removed it is an entire function of the kinematic
variables. In the physical \(s\) channel the helicity structure
imposes one more condition. The initial and final helicity
differences obey \(|\lambda-\lambda'|=4\), so rotational covariance
requires the factor \(\sin^4(\theta/2)=(1-X)^2/4\), and the
difference has the form
\begin{equation}
 \delta A_s(X)=\frac{(1-X)^2}{4}\Pi_s(X),
 \qquad \Pi_s\text{ entire},\qquad
 |\Pi_s(X)|\leq C(1+|X|)^N.
 \label{eq:treemeromorphy}
\end{equation}
Here \(C>0\) and the integer \(N\geq0\) are independent of \(X\) at
fixed \(s\), and the dependence of \(\Pi_s\) on \(s\) is suppressed
for the moment. An entire function with polynomial growth is a
polynomial, and a polynomial of degree \(N\) has a finite Jacobi
expansion. Hence \(\delta a_J^s=0\) for \(J>N+2\). In the crossed
channel, whose partial waves have helicity differences \((0,0)\), the
same reasoning gives a polynomial remainder \(\Pi_t(X_t)\) with a
finite Legendre expansion, which may still depend on \(t\). This is
the reason for matching both collinear limits rather than one. Once
both crossed operator products are imposed, the remainder is confined
to finitely many spins in each channel as a consequence of
\eqref{eq:treeclass}, and finite-spin support is no longer an
assumption about the spectrum. What remains to be shown is that
crossing and the time scaling remove this finite-spin remainder, and
that is the content of the proposition.
 
\begin{proposition}[Tree four-gluon uniqueness]
\label{prop:spectralunique}
Within the rational tree class, fix the external helicities and the
colour structure, the opposite- and same-helicity crossed light OPEs with
their cubic normalization, crossing, and a common subtraction prescription.
Impose the Carroll-time homogeneity
\begin{equation}
 \left(\sum_i u_i\partial_{u_i}+\Lambda\right)\cG^L=0.
 \label{eq:intrinsichom}
\end{equation}
Then the spectrum on positive channel invariant mass is uniquely
\eqref{eq:finitecoeff}, with the shifts fixed by the subtraction constant. The
four-point light correlator is its normalized expansion
\eqref{eq:decomposition} in the injective transform sector.
\end{proposition}

To see this, first note that \eqref{eq:blockscaling} together with
the homogeneity \eqref{eq:intrinsichom} makes the spectral
coefficients of the difference independent of \(\mu\) and
\(\nu\). The two polynomial remainders just obtained therefore have
constant coefficients, and \(\Pi_s\), \(\Pi_t\) are polynomials.
Using the same helicity phase in both channels and the angular
variables \eqref{eq:mandelstam} and \eqref{eq:tchannelangles}, the
difference reads
\begin{equation}
 \delta A_s=\frac{t^2}{s^2}\Pi_s\left(1+\frac{2t}{s}\right),
 \qquad
 \delta A_t=\Pi_t\left(1+\frac{2s}{t}\right).
 \label{eq:spectralproof}
\end{equation}
These are two descriptions of one function, so crossing requires
them to agree. With \(R=s/t\),
\begin{equation}
 R^{-2}\Pi_s(1+2/R)=\Pi_t(1+2R).
 \label{eq:laurentuniqueness}
\end{equation}
The left side is a finite sum of the powers \(R^{-2},R^{-3},\ldots\). The right side contains only nonnegative powers. Equality on an open set forces both sides to vanish, so \(\Pi_s=\Pi_t=0\). Colour recoupling multiplies by a constant finite matrix and does not affect this argument.

The physical content of the proposition is the following. The two
crossed one-particle families fix the complete two-particle tail of
the spectrum. Whatever is left is a regular tree amplitude occupying
finitely many spins, and its energy degree is incompatible with
Yang--Mills homogeneity unless it vanishes. A concrete example of
what is excluded is a contact interaction. Let \(c_8\) be a
dimensionless coefficient and \(M_*>0\) a mass scale. A contact
contribution with four powers of momentum is
\begin{equation}
 \delta\cA=\frac{c_8}{M_*^4}[13]^2\langle24\rangle^2,
 \qquad
 \delta a_J^s(\mu)=\frac{c_8\mu^2}{80\pi M_*^4}\delta_{J2},
 \qquad
 \delta b_J^t(\nu)=\frac{c_8\nu^2}{16\pi M_*^4}\delta_{J0},
 \label{eq:contactdirection}
\end{equation}
Here a fixed helicity and colour component is displayed. The remaining components are obtained by permuting the external legs as required by crossing. The Kronecker deltas select \(J=2\) in the \(s\) channel and \(J=0\) in the \(t\) channel. Thus this contribution occupies one spin in each channel. Its light transform has time weight \(\Lambda+4\), so \eqref{eq:intrinsichom} excludes it. We stress that the analyticity requirement in \eqref{eq:treemeromorphy} is imposed on complex kinematics. Regularity on the real angular interval alone would not suffice, and Appendix~\ref{app:projections} gives an explicit counterexample.

\section{Discussion and outlook}
\label{sec:discussion}

We have studied one four-point function, the alternating-helicity
four-gluon tree amplitude, from two sides. Expanding two operators
of opposite energy signs near each other gives a single exchanged
gluon with its descendants. We found that the Ward identities and the integration range (fixed by the energy signs) are sufficient to determine this contribution up to normalization. Expanding two
operators of equal energy signs gives the continuum of massive
two-particle states, and we found that the two collinear
contributions are recovered from that continuum by resummation and a
discontinuity. Once both are imposed, crossing and the time scaling
of the correlator leave no freedom in the tree spectrum except a
single subtraction constant. We now take stock of what these results rest on,
what the light basis has contributed, and where the programme goes
from here.

A few assumptions deserve to be stated once more in plain terms. The
comparison between channels uses the complete boundary data of the
light transforms on physical wave packets, and the positive-mass
pairing of Section~\ref{sec:inverse} is explicit on that space.
Contributions of zero invariant mass are not reached by this pairing
and must be supplied as separate boundary data. In the centre-of-mass
frame a real collinear configuration has zero exchanged momentum, which is
why the massless operator product had to be read off from the
analytically continued pole rather than from the real angular
interval. The convergent contour prescription defines the generic
Mellin family, and its continued Laurent coefficients retain any
boundary residues. Finally, the two-particle states we used are
Poincar\'e representations labelled by mass and spin. Rewriting them
in a basis of local composite operators at null infinity is a
separate question \cite{CotlerDhivakarJensen,CotlerEtAlQuantizing}, and it is the point of contact with the
multiparticle sectors studied in
Refs.~\cite{GuevaraHuPasterski,KulpPasterski}. The relation between
celestial locality and the Jacobi identities of soft currents has
also been studied \cite{Ball:2022bgg}.

It is worth being precise about what the light basis has and has not
added. The crossing equation \eqref{eq:crossing}, restricted to the
continuum, is helicity crossing written in a new basis, and the
bridge theorem, after resummation, is collinear factorization. The
contribution of the light basis is on the input side, in the way the
factorization data are organized into complete operator families.
Two features have no counterpart in a momentum-space pole residue.
First, the induced Ward action produces Kummer's second-order
equation before any amplitude is inserted. Its two solutions are the
descendant tower and its light-shadow, and the integration range
inherited from the energy signs fixes their relative coefficient by
cancelling the growth at large energy times time. Second, the
same-helicity block depends on the combined translation
\(v_+\partial_{u_1}+r_+\partial_{\bz_1}\). A tower built from time
descendants alone can reproduce a restricted collinear limit, but it
fails the Lorentz Ward identity. A complete celestial Mellin
transform also retains the energy information. What finite time adds
is that energy insertions and their homogeneity become explicit
inside the Ward-completed family, and this is what made the time
scaling usable as a hypothesis in the uniqueness argument.

The tree-level picture is therefore closed in the following sense.
The two crossed families determine both infinite sequences of
spectral coefficients. Once their poles are matched, rationality
confines the remaining tree contribution to finitely many spins, and
crossing with fixed Carroll-time homogeneity removes it. The common
subtraction prescription remains, and it specifies which distributional
representative of the spectrum is used. We expect this remaining
constant to be the Carrollian form of a familiar difficulty, the
forward-limit singularity that massless exchange produces in
dispersion relations, and we return to it below.

The natural next step is one loop, and the structure of the present
paper tells us what the new input is. At tree level the crossed
channels contributed poles. On the other hand, at one loop
they also contribute branch cuts, whose discontinuities provide
new input alongside the pole and subtraction data. A fixed-\(s\)
dispersion relation in the scattering angle takes the two crossed
discontinuities as input. Let the superscript \((1)\) denote the
one-loop contribution, let \(A_{\rm sub}^{(1)}\) collect the terms with no discontinuity on the open angular cuts, including subtraction terms and rational
pole contributions. Also, by \(\rho_u^{(1)}(s;\zeta)\) and
\(\rho_t^{(1)}(s;\zeta)\) let us denote the densities on the backward and
forward angular cuts. With \(\operatorname{Disc}_X\) defined as the
upper boundary value minus the lower one, their normalization for the
kernels used below is
\[
\begin{aligned}
 \rho_u^{(1)}(s;\zeta)
 &=-\frac1{2\pi\ip}
 \left.\operatorname{Disc}_X
 \left[\frac{2A^{(1)}(s,X)}{(1-X)^2}\right]\right|_{X=-\zeta},\\
 \rho_t^{(1)}(s;\zeta)
 &=-\frac1{2\pi\ip}
 \left.\operatorname{Disc}_X
 \left[\frac{2A^{(1)}(s,X)}{(1-X)^2}\right]\right|_{X=\zeta},
 \qquad \zeta>1.
\end{aligned}
\]
The subtraction terms have no discontinuity across these cuts. The
dispersion relation then reads
\begin{equation}
 A^{(1)}(s,X)=A_{\rm sub}^{(1)}(s,X)
 +\int_1^\infty\dd\zeta\left[
 \rho_u^{(1)}(s;\zeta)\frac{(1-X)^2}{2(\zeta+X)}
 -\rho_t^{(1)}(s;\zeta)\frac{(1-X)^2}{2(\zeta-X)}\right].
 \label{eq:loopdispersion}
\end{equation}
The two kernels are the displaced pole sectors of
Section~\ref{sec:partialwaves}, so the projection onto partial waves
is already known. If \(a_{J,\rm sub}^{(1)}\) denotes the Jacob-Wick
projection of \(A_{\rm sub}^{(1)}\) with the normalization of
\eqref{eq:JW}, the same pairing gives
\begin{equation}
 a_J^{(1)}(s)=a_{J,\rm sub}^{(1)}(s)
 +\frac1{8\pi}\int_1^\infty\dd\zeta\,
 \left[(-1)^J\rho_u^{(1)}\mathcal Q_{J-2}^{[4]}
       -\rho_t^{(1)}\mathcal P_{J-2}^{[4]}\right].
 \label{eq:loopinversion}
\end{equation}
This is the one-loop form of the bridge. The densities and the
subtraction terms require a common infrared factorization and
operator-mixing prescription \cite{Catani,BhardwajEtAl}.
Associativity supplies further constraints on the loop operator
product \cite{CostelloPaquette,Bittleston:2022jeq}, and an all-orders construction
exists for special twistorial theories \cite{Fernandez:2024qnu}.
Computing the crossed densities in a chosen prescription is the
immediate next calculation, and the checks in
Appendix~\ref{app:checks} test the tree identities that enter it.

Three questions become sharp at one loop. The first concerns time
scaling. Logarithms of kinematic invariants relative to a fixed
renormalization scale modify the tree homogeneity relation
\eqref{eq:intrinsichom}. Their energy dependence becomes
derivatives with respect to the Mellin weights before the light
transforms are applied. The resulting Carroll-time scaling equation
can involve Mellin derivatives of the tree correlator and mixing
between colour components. Its coefficients are controlled by the
running coupling and the soft and collinear anomalous dimensions. The second concerns the block itself. The same logarithms act
on the Tricomi combination of Section~\ref{sec:chiral}, and whether
the descendant tower and its light-shadow remain separated at one
loop, or are mixed by the infrared logarithms, is a question about the
operator content of the theory at null infinity. The third concerns
large spin. The alternating tail of Section~\ref{sec:crossing} should
acquire powers of \(\log J\) whose leading coefficient is fixed by
the one-loop crossed operator product, in close analogy with the
large-spin expansions \cite{Fitzpatrick:2012yx, Komargodski:2012ek,Caron-Huot:2017vep} of the lightcone bootstrap, and the one-loop
four-gluon amplitude is known in closed form, so this prediction can
be tested directly.

Beyond one loop, the results of this paper suggest a programme in
which the crossing equation \eqref{eq:crossing} is treated as an
equation to be solved rather than verified. Two preliminary steps
are numerical and concern the present paper itself. The light
transforms were defined as boundary values on specified contours,
and their one-to-one character on the injective sector was
established on wave packets. Both statements can be tested by
evaluating the full set of light and Mellin integrals on the actual
contours at finite Carroll times and by computing the numerical rank
of the transform, and such a test would also check the bridge
theorem at finite time rather than through its momentum-space form.
Numerical implementations of crossing and unitarity provide a
precedent for this approach \cite{Paulos:2017fhb}.
The next step is to discretize the crossing equation with unknown
spectral densities in both channels, impose the Carrollian inputs
one at a time, the pole residues, the descendant terms fixed by the Ward identities and the constraints from soft gluon currents \cite{GHPS,HPS}, and the time
scaling, and measure how much freedom remains after each. The
uniqueness proposition predicts that at tree level this freedom
shrinks to the finite-spin contact directions and then to zero. Beyond the rational tree class studied here, the numerical analysis
would also test whether subleading operator-product constraints
reduce the freedom left by crossing and the leading collinear data.

The final step is to add unitarity \cite{Lam:2017ofc}. It constrains the matrix of
partial waves between physical colour and helicity states, so the
colour-ordered coefficients must first be combined. This also
requires an infrared treatment compatible with unitarity. Allowing
higher-dimension interactions relaxes the fixed Yang-Mills time
weight and introduces local contact terms. At each fixed derivative
order these occupy finitely many spins, and their coefficients are
determined by the Wilson coefficients of those interactions.
The contact term \eqref{eq:contactdirection} gives the first example.
Crossing and unitarity can then constrain the allowed values of
these coefficients \cite{Remmen:2019cyz}. The question that
would decide whether asymptotic symmetries add anything to this
bootstrap is also quantitative. One compares the allowed region with and without the subleading constraints from the soft gluon current algebra. A smaller region would show that these constraints add information beyond the crossing and unitarity conditions imposed. An unchanged region would show that they give no additional restriction in the class tested.
In either case the common subtraction prescription of this paper
reappears as the forward-limit problem of massless dispersion
relations, and the smearing techniques developed for that problem
are the natural tool \cite{Caron-Huot:2021rmr}. Should the gluon case yield constraints, the same strategy can be
pursued for gravity using the supertranslation and stress-tensor
Ward identities. The Einstein tree amplitude has energy degree two
and hence Carroll-time weight \(\Lambda+2\). The time weights of
contact contributions depend on their own derivative order. Related
partial-wave bounds for graviton scattering have been developed in
higher dimensions \cite{Caron-Huot:2022jli}.

The present paper is the first step in this direction. It
establishes, at tree level and for one amplitude, the dictionary
between the complete operator products of light operators at null
infinity and the normalized two-particle spectrum of the amplitude.
The loop-level densities, the determinacy of the crossing equation
and the role of unitarity are the next three entries in that
dictionary.

\section*{Acknowledgements}
The research of R.~B. is partially supported by BITS Pilani
through the intramural CDRF grant C2/24/282 and by ANRF,
India, through grant ANRF/ARG/2025/009241/PS. SB acknowledges support from ANRF via CRG grant (CRG/2022/006165).

ChatGPT Astra (OpenAI) was used to assist with several
analytical steps in this work. The authors take responsibility
for the final results and conclusions.

\appendix
\section{Fusion integrals and the finite-time kernel}
\label{app:fusion}

In Section~\ref{sec:chiral} the Ward identities were derived before
the fusion integrals were evaluated. This appendix evaluates them.
It also records the energy Jacobian that enters the timelike light
blocks of Section~\ref{sec:inverse}. Since the fusion integrals and
the bridge theorem use the same integration contours and the same
branch prescriptions, we begin by fixing those.

\subsection{Light contours, branch prescriptions and the convergence domain}
\label{app:cycles}

A light transform of complex weight is an integral of a multivalued
kernel along the real line, and it is defined only once the branch of
the kernel is specified. The time prescription \eqref{eq:timeboundary}
damps each external energy integral by \(e^{-\eta_i\omega_i}\), and it
applies to both crossed blocks. For an affine light coordinate \(w\)
the integration runs from \(-\infty\) to \(+\infty\), and the branch
of the kernel is
\begin{equation}
 (r+\ip\varsigma0)^\alpha
 =\begin{cases}
 r^\alpha,&r>0,\\
 e^{\ip\pi\varsigma\alpha}|r|^\alpha,&r<0,
 \end{cases}
 \qquad \varsigma=\pm1.
 \label{eq:lightbranch}
\end{equation}
In words, the contour with \(\varsigma=+1\) passes above the branch
point of the kernel and the contour with \(\varsigma=-1\) passes below
it. Both have the same orientation. For a plus-helicity leg the
argument is \(r=w-z\) and the exponent is \(\alpha=\Delta-1\). For a
minus-helicity leg the argument is \(r=\bar w-\bz\) with the same
exponent. The light coordinate is affine, and the definition is local
on the projective light ray of the continued angular variables. When
one passes from one affine chart to another, the measure and the
field weights are transported accordingly
\cite{SharmaLight,KravchukSimmonsDuffin}.

We now specify the coordinate region in which the opposite-helicity
fusion integral is first evaluated. Choose \(z_{12}>0\), \(x>0\) and
\(W<0\). The fusion fraction runs from \(-\infty\) to \(0\), so the
factors \((-\tau)^{a-1}\) and \((1-\tau)^b\) are defined with the
real logarithm of a positive number. Write \(\varpi=\ip\eps_p\omega v\)
for the Laplace variable. The substitution \(\tau=(W/x)y\) maps the
oriented interval in \(\tau\) onto \(y\) running from \(+\infty\) to
\(0\), and the negative Jacobian reverses the limits back to the
standard integral from \(0\) to \(+\infty\). Keeping every phase
explicit, the fusion integral is
\begin{align}
 &\int_{-\infty}^{0}\dd \tau\,(-\tau)^{a-1}
 (W+\tau x+\ip\varsigma_2 0)^{b-1}e^{-\varpi\tau}\nn\\
 &\qquad=e^{-\ip\pi\varsigma_2 a}x^{-a}
 (W+\ip\varsigma_2 0)^{p-1}
 \Gamma(a)U\!\left(a,p,\frac{\varpi(W+\ip\varsigma_2 0)}x\right).
 \label{eq:orientedfusion}
\end{align}
The holomorphic light kernel of leg 1 is evaluated at the position of
leg 2 by the delta function, and in this region it contributes
\((-z_{12}+\ip\varsigma_1 0)^a=e^{\ip\pi\varsigma_1a}z_{12}^a\).
Collecting the two phases, the normalization used in
\eqref{eq:exactUblock} is
\begin{equation}
 N_g^{\varsigma_1\varsigma_2}
 =\frac g2 e^{\ip\pi(\varsigma_1-\varsigma_2)a}.
 \label{eq:cyclephase}
\end{equation}
When the two boundary signs are equal the phase cancels and
\(N_g=g/2\). For the same-helicity pair the fusion variable has the
same orientation. Putting \(y=-\tau\) gives \(y>0\) with a positive
measure, and the two kernels \((w-z_3)^{a_+}(w-z_1)^{b_+}\) keep
their individual prescriptions \eqref{eq:lightbranch} inside the
remaining \(w\)-integral. No constant phase is extracted, and the
normalization is \(N_g=g/2\), as stated in
Section~\ref{sec:plusplus}.

All the integrals used in the main text converge simultaneously in
one region of parameters and coordinates, which we call the common
convergence domain. Take generic weights with
\(\operatorname{Re}\Delta_i>1\), displaced angular poles
\(\zeta_u,\zeta_t>1\), and positive \(\eta_i\) in
\eqref{eq:timeboundary}. For the two crossed pairs of the main text
one then has
\begin{align}
 \operatorname{Re}\chi
 &=\omega(\eta_3+\eta_2)(-W/x)>0,\nn\\
 \operatorname{Re}\chi_+
 &=\omega(\eta_3+\eta_1)>0,\qquad
 \operatorname{Re}\Lambda>0,\qquad
 \operatorname{Re}Q=\eta_4+\sum_{i=1}^3Y_i\eta_i>0.
 \label{eq:commonchamber}
\end{align}
In this starting region the untransformed angular coordinate
\(\bz_1\) of the same-helicity pair and its Fourier variable \(\xi\)
are real. The momentum-conservation delta functions are imposed on
their real support first. The integration region they select is then
continued together with the light contours, and this is what
continuation of the full momentum kernel means in the main text.

For wave packets that decrease rapidly on this starting contour, the
light integrals are well defined at infinity, and the fusion and
radial integrals converge absolutely. On compact angular patches in
which the branch points stay separated, the changes of variables used
in the main text are therefore legitimate. At the two ends of the
fusion half-line the boundary term of the Ward identity behaves as
\(y^a\) at zero and is exponentially damped at infinity, so it
vanishes for \(\operatorname{Re}a>0\). The same holds with \(a_+\)
for the same-helicity pair. The displaced Jacobi sums converge
uniformly on compact physical angular intervals. Consequently
resummation, radial integration and pairing with wave packets all
agree in this domain, before the crossed pole is approached.

To reach another coordinate region, we continue the covariant
kernel, move the external points, and deform the contours
continuously, carrying the logarithms of \eqref{eq:lightbranch} along
with them. We start with the principal branch of
\(\operatorname{Log}\chi\) in \(\operatorname{Re}\chi>0\) and follow
the path without extra windings. This fixes the continued Tricomi
function together with its reflected branch. As an example, let
\(\varpi<0\) be real and let \(W\) become positive. Then
\(W+\ip\varsigma_2 0\) produces \(\chi-\ip\varsigma_2 0\) on the
negative \(\chi\) axis. The contour passes the moving root of
\(W+tx=0\) on the side inherited from the starting region, and it is
not reset to a new real contour after the substitution.

Two points about this continuation deserve emphasis. It acts on the
kernel before the kernel is paired with test functions on the new
boundary, so no analytic continuation of an arbitrary smooth test
function is ever required. And the physical timelike pairing of
Section~\ref{sec:inverse} is always taken on the Lorentzian boundary.
For tree amplitudes the continuation of the kernel is explicit. The
amplitude is a rational function of the spinor brackets
\(\langle ij\rangle\) and \([ij]\), which on the real slice are
proportional to \(z_i-z_j\) and \(\bz_i-\bz_j\). Treating \(z\) and
\(\bz\) as independent variables in these brackets gives the unique
analytic function that agrees with the amplitude on the real slice,
and this is the function on which the discontinuity
\eqref{eq:chiraldisk} acts.

The chiral discontinuity uses
\(\operatorname{Disc}(1/z)=(z+\ip0)^{-1}-(z-\ip0)^{-1}
=-2\pi\ip\delta(z)\), with the light contours held fixed on both
sides. In the bridge theorem the order of operations is as follows.
One first resums the partial waves for \(\zeta_r>1\), then takes the
meromorphic limit away from the pole, and then transports the result
to the discontinuity. Finally \(\eta_i\downarrow0\) gives the
finite-time boundary distribution. The real collinear limit is taken
with the common subtraction prescription of
Section~\ref{sec:endpointcrossing}. The principal series of weights
is reached by meromorphic continuation in the Mellin labels. At
exceptional labels, by which we mean values at which the Gamma
functions in the block have poles, the full identity is continued
first and its Laurent coefficients are taken afterwards, so that the
associated contact terms are retained.

\subsection{Fusion integrals and energy support}

We now evaluate the fusion integrals. For the opposite-helicity pair
the solution of \eqref{eq:rhoward} on the negative half-line is
\begin{equation}
 \rho(\tau)=(-\tau)^{a-1}(1-\tau)^b
 \label{eq:rhosolution}
\end{equation}
up to normalization, and the light density of \eqref{eq:unknownlight}
is therefore
\begin{equation}
 H(\tau;W,x)=(-\tau)^{a-1}(W+\tau x)^{b-1}.
 \label{eq:solvedlightdensity}
\end{equation}
In the region \(W/x<0\) we set \(\tau=(W/x)y\). After the phase of
\eqref{eq:orientedfusion} is extracted, the reduced integral is
\begin{equation}
 F(\chi)=\int_0^\infty\dd y\,y^{a-1}(1+y)^{b-1}e^{-\chi y}
 =\Gamma(a)U(a,a+b,\chi),
 \quad \operatorname{Re}a>0,\quad\operatorname{Re}\chi>0.
 \label{eq:Uint}
\end{equation}
Analytic continuation extends this to the other generic weights and
coordinate regions. The connection formula that splits the Tricomi
function into the two Kummer solutions is
\begin{equation}
 \Gamma(a)U(a,p,\chi)
 =B(a,1-p){}_1F_1(a;p;\chi)
 +\Gamma(p-1)\chi^{1-p}{}_1F_1(1-b;2-p;\chi).
 \label{eq:Uconnection}
\end{equation}
The coefficients of the analytic solution can also be obtained
directly as moments of the density,
\(\int_0^\infty\dd y\,y^{a+m-1}(1+y)^{b-1}=B(a+m,1-p-m)\), and the
identity
\begin{equation}
 \frac{B(a+m,1-p-m)}{B(a,1-p)}
 =(-1)^m\frac{(a)_m}{(p)_m}
 \label{eq:betashift}
\end{equation}
agrees with the recurrence \eqref{eq:wardrecurrence}. These moments
are understood by continuation in \(a\), because no open region of
parameters makes all of them converge at infinity at the same time.
The exact half-line integral, by contrast, converges as it stands and
retains the reflected branch.

For the same-helicity pair, the delta function in
\eqref{eq:plusparent} first sets the two holomorphic light variables
equal. Fourier transforming the remaining barred coordinate and
putting \(y=-\tau\) gives
\begin{align}
 &e^{-\ip\eps_p\omega(u_1-yv_+)}
 e^{\ip\xi(\bz_1-yr_+)}
 =e^{-\ip\eps_p\omega u_1+\ip\xi\bz_1}e^{-\chi_+y},\nn\\
 &\int_0^\infty\dd y\,y^{a_+-1}(1+y)^{b_+-1}e^{-\chi_+y}
 =\Gamma(a_+)U(a_+,a_++b_+,\chi_+).
 \label{eq:plusUint}
\end{align}
Since \(a_++b_+=p_+-1\), this is \eqref{eq:plusUblock}. The power of
\(1+y\) is fixed by the barred weight of the exchanged plus-helicity
field, and this change of weight is what shifts the second Kummer
parameter relative to the opposite-helicity case.

The boundary condition used in the Ward derivation can be checked
without solving the density equation. For an arbitrary reduced
density \(w\) on an interval \(\mathcal I_y\), put
\(F(\chi)=\int_{\mathcal I_y}\dd y\,w(y)e^{-\chi y}\). Integration by
parts gives
\begin{align}
 \mathcal D_{a,c}F={}&
 \int_{\mathcal I_y}\dd y\,e^{-\chi y}
 \{\partial_y[y(1+y)w]-(a+cy)w\}\nn\\
 &-[y(1+y)w e^{-\chi y}]_{\partial\mathcal I_y},
 \label{eq:wardboundary}
\end{align}
with \(c=p\) for the opposite-helicity pair and \(c=p_+-1\) for the
same-helicity pair. A density that satisfies the Ward equation on a
truncated interval therefore obeys Kummer's equation with a source
given by the boundary term. On the convergent contours of the
previous subsection the boundary term vanishes, and one recovers the
homogeneous equation \eqref{eq:kummerode} used in the main text.

Finally we record the energy localization used in
Section~\ref{sec:inverse}. The constraints of momentum conservation
can be solved without reference to the amplitude. At the integration
coordinates \(Z_i,\bar Z_i\) write \(q_i^\mu=q^\mu(Z_i,\bar Z_i)\),
with \(q^\mu\) as in Section~\ref{sec:conventions}. Choose three
independent rows \(\nu_1,\nu_2,\nu_3\) of the \(4\times4\) matrix
whose columns are \(\eps_i q_i^\mu\). Let \(\mathsf V\) be the
\(3\times3\) matrix formed from its first three columns and these
rows, let \(\mathsf v\) be the fourth column restricted to the same
rows, let \(\nu_4\) be the remaining row, and let
\(Y=(Y_1,Y_2,Y_3)^{\mathsf T}\) be the vector of energy ratios. Then
\begin{equation}
 Y=-\mathsf V^{-1}\mathsf v,\qquad
 \Xi=\eps_4q_4^{\nu_4}+\sum_{i=1}^3\eps_iY_iq_i^{\nu_4},\qquad
 \mathcal J_\Delta=
 \frac{(2\pi)^4\delta(\Xi)}{|\det\mathsf V|}
 \prod_{i=1}^3Y_i^{\Delta_i-1}\Theta(Y_i).
 \label{eq:jacobian}
\end{equation}
Here \(\Xi=0\) is the one remaining constraint on the angles, and
\(\Theta\) is the Heaviside step function, equal to one for positive
argument and zero for negative argument. Its value at zero plays no
role on the positive-energy wave packets used here, and any other
choice of nondegenerate rows gives the same distribution. With
\(\omega_i=Y_iE\) and \(\omega_4=E\) one has
\begin{equation}
 s=\sigma E^2,\qquad
 \sigma=4\eps_1\eps_2Y_1Y_2 Z_{12}\bar Z_{12}>0,
 \qquad
 Q=\ip\eps_4\left(u_4+
 \sum_{i=1}^3\frac{\eps_i}{\eps_4}Y_i u_i\right).
 \label{eq:energysupport}
\end{equation}
Integrating the mass delta function against the energy measure gives
\begin{equation}
 \int_0^\infty\dd E\,E^{\Lambda-1}e^{-QE}
 \delta(\sigma E^2-\mu)
 =\frac{M^{\Lambda-2}}{2\sigma^{\Lambda/2}}e^{-QM/\sqrt\sigma},
 \label{eq:radialdelta}
\end{equation}
which is the explicit mass factor in \eqref{eq:lightmasskernel}. The
remaining delta function in \(\mathcal J_\Delta\) can be used to
localize one of the four light variables. Expanding the resulting
three-fold light integral in powers of the Carroll times produces
time moments, but these are expansions of a block of fixed
\((\mu,J)\), not independent coefficients, as explained after
\eqref{eq:radialcheck}. Related angular integral representations are
given in Ref.~\cite{DeEtAl}. Generalized Euler integrals and their
differential systems provide the broader hypergeometric setting for
such integrals \cite{GKZ}, with applications to Feynman integral
representations \cite{Klausen}.

\section{Spectral pairing, angular projections and tree analyticity}
\label{app:projections}

The spectral coefficients of Section~\ref{sec:partialwaves} are
defined by a pairing, and the pairing requires an inverse of the
transform \(\mathsf T\) on a definite space of functions, together
with a common subtraction prescription at the collinear endpoints.
This appendix supplies the inverse, makes the boundary term of the
mass integration explicit, derives the closed forms and endpoint
limits of the two angular projections, and gives the counterexample
quoted in the uniqueness argument of Section~\ref{sec:spectralunique}.

\subsection{Inverting the light transform on wave packets}
\label{app:inverse}

The light transform can be undone explicitly on wave packets, and
doing so shows which boundary data must be kept for the spectral
pairing to exist. Consider one transformed angular coordinate and
write
\begin{equation}
 (\mathsf L_{\alpha,\varsigma}f)(z)
 =\int_{\mathbb R}\dd w\,(w-z+\ip\varsigma0)^\alpha f(w),
 \qquad \widetilde f(\xi)=\int_{\mathbb R}\dd z\,
 e^{-\ip\xi z}f(z).
 \label{eq:affinelight}
\end{equation}
The light transform is a convolution, so in the Fourier variable
\(\xi\) it acts by multiplication,
\begin{equation}
 \widetilde{\mathsf L_{\alpha,\varsigma}f}(\xi)
 =m_{\alpha,\varsigma}(\xi)\widetilde f(\xi),\qquad
 m_{\alpha,\varsigma}(\xi)
 =\frac{2\pi e^{\ip\pi\varsigma\alpha/2}}{\Gamma(-\alpha)}
 \Theta(-\varsigma\xi)|\xi|^{-\alpha-1}.
 \label{eq:lightmultiplier}
\end{equation}
To obtain the sign and the normalization, insert the Laplace
representation
\[
(r+\ip\varsigma0)^\alpha
=e^{\ip\pi\varsigma\alpha/2}
\int_0^\infty\dd q\,q^{-\alpha-1}e^{\ip\varsigma qr}/\Gamma(-\alpha)
\]
first for \(\operatorname{Re}\alpha<0\), where the \(q\)-integral
converges with positive damping. The \(r\)-integral then gives
\(2\pi\delta(\xi+\varsigma q)\). Continuation in \(\alpha\) gives
\eqref{eq:lightmultiplier} at generic weights, as a homogeneous
distribution \cite{GelfandShilov,DLMF}.

The step function in the multiplier is the essential point. One
boundary value of the kernel keeps only one half of the Fourier axis
and discards the other, so a single light transform is not
invertible. The two boundary values together keep both halves, and
for \(\xi\ne0\) one can reconstruct the original function from them,
\begin{equation}
 \widetilde f(\xi)=\frac{\Gamma(-\alpha)}{2\pi}
 |\xi|^{\alpha+1}
 \left[e^{-\ip\pi\alpha/2}
 \widetilde{\mathsf L_{\alpha,+}f}(\xi)
 +e^{\ip\pi\alpha/2}
 \widetilde{\mathsf L_{\alpha,-}f}(\xi)\right].
 \label{eq:lightinverse}
\end{equation}
This is why both boundary values are kept on each leg in
Section~\ref{sec:conventions}. We write \(C_c^\infty(I)\) for smooth
functions with compact support inside \(I\). A convenient dense space
of wave packets consists of functions whose angular Fourier profiles
lie in \(C_c^\infty(\mathbb R\setminus\{0\})\) and whose energy
profiles have compact support inside \((0,\infty)\). The angular
transforms and their inverses preserve this space. Its completion in
the Hilbert norm restores arbitrarily small nonzero frequencies, and
no nonzero square-integrable packet is concentrated at a single
frequency, so nothing is lost by excluding \(\xi=0\).

The modified Mellin transform is inverted at fixed generic
\(\Delta\). With \(O_{\Delta,\ell}^{\eps}(u-\ip\eps\eta)\) the field
at the regulated time of \eqref{eq:timeboundary},
\begin{equation}
 a_\ell^\eps(\omega)
 =e^{\eta\omega}\omega^{1-\Delta}
 \int_{\mathbb R}\frac{\dd u}{2\pi}\,
 e^{\ip\eps\omega u}
 O_{\Delta,\ell}^{\eps}(u-\ip\eps\eta),\qquad \omega>0.
 \label{eq:mminverse}
\end{equation}
The inverse of the complete transform \(\mathsf T\) of
Section~\ref{sec:conventions} is now defined as the composition of
these two steps. One first applies \eqref{eq:lightinverse} to each of
the four legs, using both boundary values of its light kernel, and
then applies \eqref{eq:mminverse} to each leg. This reconstructs the
momentum kernel on the image of \(\mathsf T\), and it is the operator
\(\mathsf T^{-1}\) used in Section~\ref{sec:crossing}. It is defined
on that image, whose boundary components are not independent but
satisfy the compatibility implied by their common preimage.
Scattering amplitudes and fixed-mass partial waves enter this
construction as distributional limits of packet matrix elements, and
their continuation follows Appendix~\ref{app:cycles}. At exceptional
weights the generic meromorphic family is inverted first and its
limit or residue is taken afterwards.

For the pairing we use physical helicity states and the spaces
\begin{align}
 \mathscr H_{\rm red}
 &=L^2\!\left((0,\infty)\times[-1,1],\dd s\,\dd X\right),\nn\\
 \mathscr D_{\rm red}
 &=\left\{\sum_{J=2}^{J_*} f_J(s)d_J(X)\,,\quad
 f_J\in C_c^\infty(0,\infty)\right\}.
 \label{eq:packetdomain}
\end{align}
Here \(L^2\) denotes square-integrable functions in the measure
shown, and \(J_*\) is any finite integer at least two. By
completeness of the Jacobi polynomials, \(\mathscr D_{\rm red}\) is
dense in \(\mathscr H_{\rm red}\). For distributional spectra one also
admits rapidly decreasing spin coefficients with a common compact mass
support and the same decay for all mass derivatives. This test space
supports the pairing with the regulated pole spectra and with their
finite-part limits. Smooth packets in the overall boost and
orientation variables are understood before passing to the reduced
kernel. The transform of a finite sum in \(\mathscr D_{\rm red}\) is
the corresponding mass packet of the blocks \eqref{eq:massblock}. The
Fourier construction above justifies its inverse, and the completion
fixes its extension. Analytic continuation is always performed on the
covariant kernels before packet matrix elements are taken, as in
Appendix~\ref{app:cycles}.

In the pairing formulas below, \(\mathsf T F\) abbreviates the
transform of the full momentum kernel,
\[
 \mathsf T\!\left[
 (2\pi)^4\delta^{(4)}\!\left(\sum_i p_i\right)
 \Phi f(s,X)\right].
\]
Let \(\mathscr H_L\) be the completion of
\(\mathsf T\mathscr D_{\rm red}\) in the norm defined by
\begin{equation}
 (\mathsf T f_1,\mathsf T f_2)_L=(f_1,f_2)_{\rm red}.
 \label{eq:pullbackdomain}
\end{equation}
Because the inverse is explicit, this pairing does not depend on the
choice of representative once the complete boundary data are used.
Gauge null states have already been removed by working with physical
helicities, and at generic weights the complete transform has no
further null space on wave packets. If one restricts to a smaller
set of contours, the restricted transform acquires a kernel. In that
case let \(\mathscr N\) be the closure in the packet norm of
\(\ker\mathsf T_{\rm restricted}\), and define the pairing on the
quotient by using the orthogonal representative in
\(\mathscr N^\perp\). The Gram matrix is then that of the projected
partial waves. Equation \eqref{eq:gram} refers to the complete data,
for which every Jacob--Wick wave is retained, so a quotient never
assigns a norm to a discarded component.

The pole amplitudes are not elements of \(\mathscr H_{\rm red}\) but
generalized vectors, and their coefficients are obtained by pairing
with the test space above, with the chosen subtraction prescription.
The distributional meaning of \eqref{eq:inversion} is
\begin{equation}
 \langle a_J,\varphi\rangle
 =-\frac1{8\pi}
 \left\langle \cR(\mu,X),\frac{\varphi(\mu)}\mu d_J(X)\right\rangle,
 \qquad \varphi\in C_c^\infty(0,\infty).
 \label{eq:weakinversion}
\end{equation}
Multiplication by \(1/\mu\) is smooth on the support of \(\varphi\),
so the Casimir source determines the positive-mass spectral
distribution uniquely. A contribution supported at \(\mu=0\) is not
reached, because multiplication by \(\mu\) annihilates
\(\delta(\mu)\). Such contributions are the separate boundary data
referred to in Section~\ref{sec:inverse}.

\subsection{Mass integration and time homogeneity}
\label{app:massends}

The scaling argument of Section~\ref{sec:inverse} integrates by parts
in the mass, and here we keep the boundary term of that integration
explicit. For a finite mass interval and
\(\mathcal E_u=\sum_i u_i\partial_{u_i}\), the block scaling
\eqref{eq:blockscaling} gives
\begin{align}
 (\mathcal E_u+\Lambda+k)
 \int_{\mu_-}^{\mu_+}\dd\mu\,a_J\cB_{\mu J}^{L,s}
 ={}&2\left[\mu a_J\cB_{\mu J}^{L,s}\right]_{\mu_-}^{\mu_+}
 \nn\\
 &+\int_{\mu_-}^{\mu_+}\dd\mu\,
 (k a_J-2\mu\partial_\mu a_J)\cB_{\mu J}^{L,s}.
 \label{eq:massboundary}
\end{align}
For a mass packet of compact support the boundary term vanishes
exactly, which proves the relation \(2\mu\partial_\mu a_J=ka_J\) on
\((0,\infty)\) in the weak sense. For the homogeneous tree density
the direct radial integral has the same property in the starting
domain. The radial factor of \(\cB_{\mu J}\) is
\(\mu^{(\Lambda-2)/2}e^{-Q\sqrt{\mu/\sigma}}\), so for
\(a_J\propto\mu^{k/2}\) one has
\begin{equation}
 \mu a_J\cB_{\mu J}
 =O\!\left(\mu^{(\Lambda+k)/2}\right)
 \quad(\mu\downarrow0),
 \qquad \operatorname{Re}(\Lambda+k)>0,
 \label{eq:massendpointfalloff}
\end{equation}
while \(\operatorname{Re}Q>0\) controls the behaviour at infinity. In
this scaling the complex times, including their imaginary parts,
scale together. The relation at physical times follows by
continuation of the identity. At a Mellin pole the identity holds
between meromorphic families, and its Laurent coefficients are taken
together, so that any boundary residue and the corresponding term in
the scaling equation of the finite part are retained. The uniqueness
statement at positive mass is unaffected by this, and any extension
to zero mass is specified separately.

\subsection{The regulated angular projections}

For \(n=J-2\), the projection functions \(\mathcal Q_n^{[4]}\) and
\(\mathcal P_n^{[4]}\) are defined in \eqref{eq:QPdefinitions}.
Rodrigues' formula for the Jacobi polynomial, followed by \(n\)
integrations by parts, moves all derivatives onto the displaced pole
and gives the representations
\begin{align}
 \mathcal Q_n^{[4]}(\zeta)
 &=2^n\int_0^1
 \frac{v^{n+4}(1-v)^n}{(\zeta+1-2v)^{n+1}}\dd v,\nn\\
 \mathcal P_n^{[4]}(\zeta)
 &=2^n\int_0^1
 \frac{v^n(1-v)^{n+4}}{(\zeta+1-2v)^{n+1}}\dd v,
 \label{eq:positiveQP}
\end{align}
whose integrands are positive for \(\zeta>1\). These are the forms
used at large spin in Appendix~\ref{app:checks}, since they involve
no cancellations. Expanding the integrands gives the hypergeometric
forms
\begin{align}
 \mathcal Q_n^{[4]}(\zeta)
 &=\frac{2^n B(n+1,n+5)}{(\zeta+1)^{n+1}}
 {}_2F_1\!\left(n+1,n+5;2n+6;\frac2{\zeta+1}\right),\nn\\
 \mathcal P_n^{[4]}(\zeta)
 &=\frac{2^n B(n+1,n+5)}{(\zeta+1)^{n+1}}
 {}_2F_1\!\left(n+1,n+1;2n+6;\frac2{\zeta+1}\right).
 \label{eq:QPclosed}
\end{align}
Here \({}_2F_1\) is the Gauss hypergeometric function, normalized for
\(|z|<1\) by
\[
 {}_2F_1(a,b;c;z)
 =\sum_{m=0}^{\infty}\frac{(a)_m(b)_m}{(c)_m\,m!}z^m
\]
and continued analytically to other arguments. The first of these
is a Froissart--Gribov projection with Jacobi weight. It can also be
written in terms of the ordinary Legendre functions of the second
kind. Expand the polynomial
\((1+y)^4P_n^{(0,4)}(y)=\sum_{k=0}^{n+4}q_{nk}P_k(y)\). Then
\begin{equation}
 \mathcal Q_n^{[4]}(\zeta)=\frac1{16}
 \sum_{k=0}^{n+4}q_{nk}Q_k(\zeta),\qquad
 q_{nk}=\frac{2k+1}{2}\int_{-1}^1\dd y\,
 (1+y)^4P_n^{(0,4)}(y)P_k(y).
 \label{eq:LegendreQ}
\end{equation}
The same expansion of the forward numerator expresses
\(\mathcal P_n^{[4]}\) in the same basis of \(Q_k\).

We now take the limits that remove the regulators. In the first
line of \eqref{eq:QPclosed} the third parameter of the hypergeometric
function equals the sum of the first two, and in that case the
function has a logarithmic singularity as its argument approaches
one. With \(\psi(z)=\Gamma'(z)/\Gamma(z)\) the digamma function, the
expansion is
\begin{equation}
 \mathcal Q_n^{[4]}(1+\delta)
 =\frac12\left[\log\frac2\delta
 +2\psi(1)-\psi(n+1)-\psi(n+5)\right]+o(1).
 \label{eq:Qendpoint}
\end{equation}
Using \(\psi(m+1)=H_m-\gamma_E\) turns the finite part into the
harmonic numbers of \eqref{eq:Qlimits}. In the second line of
\eqref{eq:QPclosed} the parameters are such that Gauss' summation
formula applies directly at unit argument, and one finds
\begin{equation}
 \mathcal P_n^{[4]}(1)
 =\frac{3\Gamma(n+1)}{\Gamma(n+5)}
 =\frac3{(n+1)(n+2)(n+3)(n+4)}.
 \label{eq:Pendpoint}
\end{equation}
Equivalently, at \(\zeta=1\) the positive integral in
\eqref{eq:positiveQP} reduces to \(\tfrac12 B(n+1,4)\). The forward
projection is therefore finite and needs no subtraction in this
helicity component, as stated in Section~\ref{sec:partialwaves}.
Together, \eqref{eq:Qendpoint} and \eqref{eq:Pendpoint} establish the
two sequences in \eqref{eq:finitecoeff}.

\subsection{Tree analyticity and finite spin}

The uniqueness argument of Section~\ref{sec:spectralunique} uses
analyticity on complex kinematics, not merely regularity on the real
angular interval. Here we give the example showing that the weaker
condition would not suffice. Let \(\gamma\) be a dimensionless
constant and let \(\bm\cA_{\rm YM}\) be the colour-dressed Yang--Mills
tree amplitude of \eqref{eq:DDM}. The deformation
\begin{equation}
 \delta\bm\cA
 =\gamma\bm\cA_{\rm YM}
 \frac{s^2t^2u_{\!M}^2}{(s^2+t^2+u_{\!M}^2)^3}
 \label{eq:nonlocalcounter}
\end{equation}
has degree zero in the energies and leaves every physical pole
residue unchanged, because the multiplying factor vanishes on each
collinear locus. Its \(1234\) angular component is
\begin{equation}
 \frac{\delta A_{1234}}{g^2\gamma}
 =\frac{(1-X)^4(1+X)}{4(X^2+3)^3}.
 \label{eq:counterangular}
\end{equation}
This function is regular at both the forward and the backward
collinear limits, so it would pass any test formulated on the real
interval. It has complex poles at \(X=\pm\ip\sqrt3\), however, and
therefore an infinite Jacobi expansion. Its coefficients decay
exponentially in the spin, as expected for a function analytic in a
finite Bernstein ellipse \cite{JacobiDecay}, so even a decreasing
bound on the large-spin coefficients would not exclude it. What
excludes it is the absence of complex poles together with polynomial
growth at infinity. These two properties are what turn the
pole-subtracted helicity function into a polynomial in
\eqref{eq:treemeromorphy}, and they are the assumptions under which
finite-spin support is derived.

\section{Numerical checks}
\label{app:checks}
The python scripts for the numerical checks are available \href{https://drive.google.com/drive/folders/1VCE0QmWxcf-3Jw75sTfMDiQ-HY27k2Cv?usp=drive_link}{here}, and will be shared upon request.

The script \texttt{verify\_decomposition.py} checks the
normalizations, the Ward operators and the two channel bridges using
independent projections and quadrature. It requires Python, NumPy and
SciPy, and all numbers below use \(g^2=1\).

The Lorentz Ward identities of the parent kernels are tested on 96
arbitrary monomial fields for each helicity pair in exact rational
arithmetic, and the induced light action for the opposite-helicity
pair is checked on 36 configurations with arbitrary densities. All of
these identities hold exactly. The Ward identity in the Laplace
variable, \eqref{eq:wardlaplace}, is also checked on densities that
do not solve the Ward equation and on truncated contours, where the
boundary term of \eqref{eq:wardboundary} is nonzero and is reproduced.

\begin{center}
\begin{tabular}{lr}
\toprule
Check & Maximum error\\
\midrule
Angular orthogonality & \(2.4\times10^{-16}\)\\
Both regulated pole projections & \(1.1\times10^{-13}\)\\
Finite-part projections & \(2.2\times10^{-15}\)\\
Forward pole, direct projection & \(2.7\times10^{-18}\)\\
Ward--Laplace operator identity, scaled & \(5.4\times10^{-16}\)\\
Opposite-helicity Tricomi integral, relative & \(1.8\times10^{-12}\)\\
Same-helicity Tricomi integral, relative & \(6.5\times10^{-13}\)\\
Same-helicity light wave packet, relative & \(9.1\times10^{-14}\)\\
Same-helicity Mellin measure and phase & \(7.1\times10^{-16}\)\\
Adjacent-spin identity & \(7.5\times10^{-17}\)\\
Polynomial contact, spins above its support & \(1.4\times10^{-18}\)\\
\bottomrule
\end{tabular}
\end{center}

The same-helicity wave-packet test keeps both \(v_+\) and \(r_+\)
nonzero, and direct integration over the fusion fraction and the
light variable agrees with the reduced Tricomi kernel. This check is
therefore sensitive to the angular term in \(\chi_+\). An independent
spinor test of the collinear limit \eqref{eq:plussplitting} gives
relative errors \(3.40\times10^{-3}\), \(3.40\times10^{-5}\) and
\(3.40\times10^{-7}\) at collinear deformations \(10^{-3}\),
\(10^{-5}\) and \(10^{-7}\), consistent with a linear approach.

For the regulated reconstruction of the two pole sectors, both
propagator poles are displaced with \(\zeta=1.15\) and the partial
sums are evaluated at \(X=-0.8,-0.4,0,0.3,0.7,0.9\). At
\(J_{\max}=48\) both ordered partial sums agree with their closed
forms to \(2.9\times10^{-11}\), and at \(J_{\max}=96\) the maximum
error is \(3.6\times10^{-15}\). The positive representations
\eqref{eq:positiveQP} are used at large spin to avoid the
cancellations present in the hypergeometric forms.

The unregulated forward contribution converges only algebraically.
On the same angular set with \(X=0.98\) added, its maximum absolute
errors at \(J_{\max}=32,128,512\) are \(1.54\times10^{-4}\),
\(1.06\times10^{-6}\) and \(3.50\times10^{-8}\). At the last
truncation \(-2A_t(0.98)/(1-0.98)=1.000003501\), which approaches the
expected forward residue. The JSON file written by the script also
records the reconstruction near the backward collinear limit, both
physical-channel projections, the radial and Casimir-weighted mass
integrals of \eqref{eq:radialcheck}, the test of the subtraction term
in \eqref{eq:s_endpointfamily}, and the two contact projections of
\eqref{eq:contactdirection}.

A second script, \texttt{verify\_cycles\_and\_domain.py}, checks the
prescriptions of Appendices~\ref{app:cycles} and \ref{app:inverse}.
Direct quadrature of the opposite-helicity fusion integral agrees
with the continued Tricomi expression on both signs of \(W\) and for
all four choices of boundary signs, with maximum relative error
\(7.8\times10^{-13}\). Angular convolution on packets built from
derivatives of Gaussians checks the sign and normalization of the
Fourier multiplier \eqref{eq:lightmultiplier} to \(9.6\times10^{-12}\).
The mass integration-by-parts identity \eqref{eq:massboundary} agrees
to \(1.7\times10^{-15}\), both for compact packets and for a finite
window with a nonzero boundary term. These are independent checks of
the local formulas. The continuation of the full bridge relies on the
common-domain argument of Appendix~\ref{app:cycles}.


\bibliographystyle{unsrt}
\begingroup
\let\originalthebibliography\thebibliography
\renewcommand{\thebibliography}[1]{%
  \originalthebibliography{#1}%
  \small
  \setlength{\itemsep}{1pt}%
  \setlength{\parskip}{0pt}%
}
\bibliography{light_bootstrap_references}
\endgroup
\end{document}